\documentclass[aps,pra,10pt,tightenlines,amsmath,amssymb,amsfonts,twocolumn,floatfix,nofootinbib,showpacs]{revtex4-1}

\usepackage{hyperref}
\usepackage[figure]{hypcap}
\usepackage{amsthm}
\usepackage{graphicx}
\usepackage{color}
\usepackage{bm}
\usepackage{times}
\usepackage{mathtools}
\usepackage{ifthen}
\usepackage{enumerate}
\usepackage{array}
\usepackage{tabularx}
\usepackage{qcircuit}
\usepackage{pgfmath}

\usepackage{tikz}
\usetikzlibrary{chains,calc,fit,shapes.geometric,decorations.pathmorphing}
\pgfrealjobname{Temporal_offline}

\DeclareMathOperator{\sech}{sech}

\def\tp{\mathrm{T}}

\def\op#1{\hat{#1}}
\def\opvec#1{\op{\vec{#1}}}

\def\id{I}

\def\1{\mat{\id}}

\def\mat#1{\bm{\mathrm{#1}}}

\renewcommand{\vec}[1]{\bm{\mathrm{#1}}}

\def\controlled#1{\mathrm{C}_{#1}}
\def\CZ{\controlled Z}

\def\cH{\mathcal{H}}
\def\cC{\mathcal{C}}

\def\micronodesize{4pt}
\def\edgethickness{very thick}
\def\poscolor{blue!60!black}
\def\negcolor{red!30!yellow}

\def\unitcolor{\poscolor}
\def\evencolor{black}
\def\oddcolor{white}
\def\phantomcolor{gray!90}
\def\phantomfade{very nearly transparent}

\tikzset{>=to}

\tikzset{micro-no-color/.style={
	circle,
	minimum size=\micronodesize,
	inner sep=0pt
	}
}

\tikzset{micro/.style={
	micro-no-color, ball color=black,
	}
}

\tikzset{micro-even/.style={
	micro-no-color, ball color=\evencolor,
	}
}

\tikzset{micro-odd/.style={
	micro-no-color, ball color=\oddcolor,
	}
}

\tikzset{poslink/.style={
	\edgethickness,
	draw=\poscolor,
	nearly opaque
	}
}

\tikzset{neglink/.style={
	\edgethickness,
	draw=\negcolor,
	nearly opaque
	}
}

\tikzset{unitlink/.style={
	\edgethickness,
	draw=\unitcolor,
	nearly opaque
	}
}

\tikzset{dashlink/.style={
	\edgethickness,
	draw=\unitcolor,
	dash pattern=on 0.75pt off 0.75pt,
	nearly opaque
	}
}

\tikzset{fulllink/.style={
	\edgethickness,
	draw=\unitcolor,
	opaque
	}
}

\tikzset{phantomlink/.style={
	\edgethickness,
	draw=\phantomcolor,
	\phantomfade
	}
}

\tikzset{optpath/.style={
	thick,
	draw=black
	}
}

\tikzset{optdelay/.style={
	optpath,
	decorate,
	decoration={coil,segment length=4pt,pre=lineto,pre length=1mm,post length=1mm}
	}
}

\tikzset{optdelaylong/.style={
	optpath,
	decorate,
	decoration={coil,segment length=2pt, pre=lineto, pre length=1mm,post length=1mm}
	}
}

\tikzset{nodehighlight/.style={
	semithick,
	red,
	fill=red,
	semitransparent
	}
}

\def\latticeoptsscale{0.8}

\tikzset{latticeopts/.style={
	baseline=-\latticeoptsscale*1.5cm-1ex,
	x= \latticeoptsscale*3.375cm,
	y= \latticeoptsscale*3.75mm,
	z= \latticeoptsscale*20.25mm,
	inner sep=0pt,
	outer sep=0pt
	}
}
	
\newcommand\piplus[4] 
{
	\draw [poslink] (#1) -- (#3);
	\draw [poslink] (#1) -- (#4);
	\draw [poslink] (#2) -- (#3);
	\draw [poslink] (#2) -- (#4);	
}

\newcommand\piminus[4] 
{
	\draw [poslink] (#1) -- (#3);
	\draw [neglink] (#1) -- (#4);
	\draw [neglink] (#2) -- (#3);
	\draw [poslink] (#2) -- (#4);	
}

\newcommand\beamsplit[4] 
{
	\draw [neglink] (#1) -- (#3);
	\draw [neglink] (#1) -- (#4);
	\draw [poslink] (#2) -- (#3);
	\draw [poslink] (#2) -- (#4);	
}

\newcommand\tmslink[4] 
{
	\draw [phantomlink] (#1) -- (#3);	
	\draw [phantomlink] (#1) -- (#4);	
	\draw [unitlink] (#2) -- (#3);	
	\draw [phantomlink] (#2) -- (#4);	
}

\newcommand\linearlink[4] 
{
	\draw [phantomlink] (#1) -- (#3);	
	\draw [phantomlink] (#1) -- (#4);	
	\draw [phantomlink] (#2) -- (#3);	
	\draw [unitlink] (#2) -- (#4);	
}

\newcommand\Pizero[8] 
{
	\draw [poslink] (#1) -- (#5);
	\draw [poslink] (#1) -- (#6);
	\draw [poslink] (#1) -- (#7);
	\draw [poslink] (#1) -- (#8);
	\draw [poslink] (#2) -- (#5);
	\draw [poslink] (#2) -- (#6);	
	\draw [poslink] (#2) -- (#7);
	\draw [poslink] (#2) -- (#8);	
	\draw [poslink] (#3) -- (#5);
	\draw [poslink] (#3) -- (#6);
	\draw [poslink] (#3) -- (#7);
	\draw [poslink] (#3) -- (#8);
	\draw [poslink] (#4) -- (#5);
	\draw [poslink] (#4) -- (#6);	
	\draw [poslink] (#4) -- (#7);
	\draw [poslink] (#4) -- (#8);	
}

\newcommand\Pione[8] 
{
	\draw [poslink] (#1) -- (#5);
	\draw [neglink] (#1) -- (#6);
	\draw [poslink] (#1) -- (#7);
	\draw [neglink] (#1) -- (#8);
	\draw [neglink] (#2) -- (#5);
	\draw [poslink] (#2) -- (#6);	
	\draw [neglink] (#2) -- (#7);
	\draw [poslink] (#2) -- (#8);	
	\draw [poslink] (#3) -- (#5);
	\draw [neglink] (#3) -- (#6);
	\draw [poslink] (#3) -- (#7);
	\draw [neglink] (#3) -- (#8);
	\draw [neglink] (#4) -- (#5);
	\draw [poslink] (#4) -- (#6);	
	\draw [neglink] (#4) -- (#7);
	\draw [poslink] (#4) -- (#8);	
}

\newcommand\Pitwo[8] 
{
	\draw [poslink] (#1) -- (#5);
	\draw [poslink] (#1) -- (#6);
	\draw [neglink] (#1) -- (#7);
	\draw [neglink] (#1) -- (#8);
	\draw [poslink] (#2) -- (#5);
	\draw [poslink] (#2) -- (#6);	
	\draw [neglink] (#2) -- (#7);
	\draw [neglink] (#2) -- (#8);	
	\draw [neglink] (#3) -- (#5);
	\draw [neglink] (#3) -- (#6);
	\draw [poslink] (#3) -- (#7);
	\draw [poslink] (#3) -- (#8);
	\draw [neglink] (#4) -- (#5);
	\draw [neglink] (#4) -- (#6);	
	\draw [poslink] (#4) -- (#7);
	\draw [poslink] (#4) -- (#8);	
}

\newcommand\Pithree[8] 
{
	\draw [poslink] (#1) -- (#5);
	\draw [neglink] (#1) -- (#6);
	\draw [neglink] (#1) -- (#7);
	\draw [poslink] (#1) -- (#8);
	\draw [neglink] (#2) -- (#5);
	\draw [poslink] (#2) -- (#6);	
	\draw [poslink] (#2) -- (#7);
	\draw [neglink] (#2) -- (#8);	
	\draw [neglink] (#3) -- (#5);
	\draw [poslink] (#3) -- (#6);
	\draw [poslink] (#3) -- (#7);
	\draw [neglink] (#3) -- (#8);
	\draw [poslink] (#4) -- (#5);
	\draw [neglink] (#4) -- (#6);	
	\draw [neglink] (#4) -- (#7);
	\draw [poslink] (#4) -- (#8);	
}

\newcommand\BeamSplitx[8] 
{
	\draw [neglink] (#1) -- (#5);
	\draw [neglink] (#1) -- (#6);
	\draw [neglink] (#1) -- (#7);
	\draw [neglink] (#1) -- (#8);
	\draw [poslink] (#2) -- (#5);
	\draw [poslink] (#2) -- (#6);	
	\draw [poslink] (#2) -- (#7);
	\draw [poslink] (#2) -- (#8);	
	\draw [neglink] (#3) -- (#5);
	\draw [neglink] (#3) -- (#6);
	\draw [neglink] (#3) -- (#7);
	\draw [neglink] (#3) -- (#8);
	\draw [poslink] (#4) -- (#5);
	\draw [poslink] (#4) -- (#6);	
	\draw [poslink] (#4) -- (#7);
	\draw [poslink] (#4) -- (#8);	
}

\newcommand\BeamSplitz[8] 
{
	\draw [neglink] (#1) -- (#5);
	\draw [neglink] (#1) -- (#6);
	\draw [poslink] (#1) -- (#7);
	\draw [poslink] (#1) -- (#8);
	\draw [poslink] (#2) -- (#5);
	\draw [poslink] (#2) -- (#6);	
	\draw [neglink] (#2) -- (#7);
	\draw [neglink] (#2) -- (#8);	
	\draw [poslink] (#3) -- (#5);
	\draw [poslink] (#3) -- (#6);
	\draw [neglink] (#3) -- (#7);
	\draw [neglink] (#3) -- (#8);
	\draw [neglink] (#4) -- (#5);
	\draw [neglink] (#4) -- (#6);	
	\draw [poslink] (#4) -- (#7);
	\draw [poslink] (#4) -- (#8);	
}

\newcommand\tmslinkx[8] 
{
	\draw [phantomlink] (#1) -- (#5);
	\draw [phantomlink] (#1) -- (#6);
	\draw [phantomlink] (#1) -- (#7);
	\draw [phantomlink] (#1) -- (#8);
	\draw [unitlink] (#2) -- (#5);
	\draw [phantomlink] (#2) -- (#6);	
	\draw [phantomlink] (#2) -- (#7);
	\draw [phantomlink] (#2) -- (#8);	
	\draw [phantomlink] (#3) -- (#5);
	\draw [phantomlink] (#3) -- (#6);
	\draw [phantomlink] (#3) -- (#7);
	\draw [phantomlink] (#3) -- (#8);
	\draw [phantomlink] (#4) -- (#5);
	\draw [phantomlink] (#4) -- (#6);	
	\draw [phantomlink] (#4) -- (#7);
	\draw [phantomlink] (#4) -- (#8);
}

\newcommand\tmslinkz[8] 
{
	\draw [phantomlink] (#1) -- (#5);
	\draw [phantomlink] (#1) -- (#6);
	\draw [phantomlink] (#1) -- (#7);
	\draw [phantomlink] (#1) -- (#8);
	\draw [phantomlink] (#2) -- (#5);
	\draw [phantomlink] (#2) -- (#6);	
	\draw [phantomlink] (#2) -- (#7);
	\draw [phantomlink] (#2) -- (#8);	
	\draw [phantomlink] (#3) -- (#5);
	\draw [phantomlink] (#3) -- (#6);
	\draw [phantomlink] (#3) -- (#7);
	\draw [phantomlink] (#3) -- (#8);
	\draw [phantomlink] (#4) -- (#5);
	\draw [phantomlink] (#4) -- (#6);	
	\draw [unitlink] (#4) -- (#7);
	\draw [phantomlink] (#4) -- (#8);
}

\newcommand\linearlinklattice[8] 
{
	\draw [phantomlink] (#1) -- (#5);
	\draw [phantomlink] (#1) -- (#6);
	\draw [phantomlink] (#1) -- (#7);
	\draw [phantomlink] (#1) -- (#8);
	\draw [phantomlink] (#2) -- (#5);
	\draw [phantomlink] (#2) -- (#6);	
	\draw [phantomlink] (#2) -- (#7);
	\draw [phantomlink] (#2) -- (#8);	
	\draw [phantomlink] (#3) -- (#5);
	\draw [phantomlink] (#3) -- (#6);
	\draw [phantomlink] (#3) -- (#7);
	\draw [phantomlink] (#3) -- (#8);
	\draw [phantomlink] (#4) -- (#5);
	\draw [phantomlink] (#4) -- (#6);	
	\draw [phantomlink] (#4) -- (#7);
	\draw [unitlink] (#4) -- (#8);
}

\newcommand\beamsplitx[8] 
{
	\draw [neglink] (#1) -- (#5);
	\draw [neglink] (#1) -- (#6);
	\draw [phantomlink] (#1) -- (#7);
	\draw [phantomlink] (#1) -- (#8);
	\draw [poslink] (#2) -- (#5);
	\draw [poslink] (#2) -- (#6);	
	\draw [phantomlink] (#2) -- (#7);
	\draw [phantomlink] (#2) -- (#8);	
	\draw [phantomlink] (#3) -- (#5);
	\draw [phantomlink] (#3) -- (#6);
	\draw [phantomlink] (#3) -- (#7);
	\draw [phantomlink] (#3) -- (#8);
	\draw [phantomlink] (#4) -- (#5);
	\draw [phantomlink] (#4) -- (#6);	
	\draw [phantomlink] (#4) -- (#7);
	\draw [phantomlink] (#4) -- (#8);
}

\newcommand\beamsplitz[8] 
{
	\draw [phantomlink] (#1) -- (#5);
	\draw [phantomlink] (#1) -- (#6);
	\draw [phantomlink] (#1) -- (#7);
	\draw [phantomlink] (#1) -- (#8);
	\draw [phantomlink] (#2) -- (#5);
	\draw [phantomlink] (#2) -- (#6);	
	\draw [phantomlink] (#2) -- (#7);
	\draw [phantomlink] (#2) -- (#8);	
	\draw [phantomlink] (#3) -- (#5);
	\draw [phantomlink] (#3) -- (#6);
	\draw [neglink] (#3) -- (#7);
	\draw [neglink] (#3) -- (#8);
	\draw [phantomlink] (#4) -- (#5);
	\draw [phantomlink] (#4) -- (#6);	
	\draw [poslink] (#4) -- (#7);
	\draw [poslink] (#4) -- (#8);
}

\newcommand\bs[3] 
{
	\draw [thin,fill=white,semitransparent] #1 +(-0.5*#2,-0.5*#3) rectangle +(0.5*#2,0.5*#3);
}

\newcommand\squeezedstate[4] 
{
	\draw [thin,gray] #1 +(-0.5*#2,0) -- +(0.5*#2,0) +(0,-0.5*#2) -- +(0,0.5*#2);
	\draw [thin] #1 circle (#3 and #4);
}

\begin{document}

\title{Temporal-mode continuous-variable cluster states using linear optics}

\author{Nicolas C. Menicucci}
\affiliation{Perimeter Institute for Theoretical Physics, Waterloo, Ontario N2L 2Y5, Canada}

\begin{abstract}
I present an extensible experimental design for optical continuous-variable cluster states of arbitrary size using four offline (vacuum) squeezers and six beamsplitters.  This method has all the advantages of a temporal-mode encoding [Phys.~Rev.~Lett.\ \textbf{104},~250503], including finite requirements for coherence and stability even as the computation length increases indefinitely, with none of the difficulty of inline squeezing.  The extensibility stems from a construction based on Gaussian projected entangled pair states~(GPEPS).  The potential for use of this design within a fully fault tolerant model is discussed.
\end{abstract}

\date{July 20, 2010}


\maketitle

\section{Introduction}%
\label{sec:intro}

Quantum computation~(QC) is the controlled coherent manipulation of quantum information.  Analogous to its classical counterpart, quantum information is usually encoded in the quantum state of locally addressable systems, and its manipulation is performed by inducing coherent unitary evolution of these systems via external laboratory equipment~\cite{Nielsen2000}.  The one-way model of QC~\cite{Raussendorf2001} replaces this need for coherent unitary control by a sequence of adaptive local measurements made on a highly entangled resource state called a \emph{cluster state}~\cite{Briegel2001}.  This resource acts as a universal substrate with quantum information encoded virtually within it.  With local projective measurements often being easier to implement than unitary evolution, the philosophy of the model is this: perfect the creation of cluster states, and the rest is (comparatively) easy.

In its continuous-variable (CV) incarnation, universal one-way QC~\cite{Menicucci2006} requires a resource state known as a \emph{continuous variable cluster state}~\cite{Zhang2006}, which is a multimode squeezed Gaussian state.  In an optical setting, homodyne detection and photon counting---plus classical feedforward of measurement results---suffice to implement universal QC using CVs~\cite{Gu2009}.  Homodyne detection alone is sufficient to implement all multimode Gaussian operations~\cite{Gu2009}, given a cluster state with a sufficiently connected graph.\footnote{While usage varies in the literature, I am using the convention~\cite{Nielsen2006,Menicucci2006,Menicucci2008,Flammia2009,Menicucci2010,Menicucci2010} that a ``cluster state'' can have any associated graph.  Some authors prefer to call this a ``graph state,'' reserving ``cluster state'' for square-lattice graph states only.  I would refer to such states as ``cluster states with a square-lattice graph.''}

There are currently four proposed methods for constructing universal CV cluster states optically~\cite{Menicucci2006,vanLoock2007,Flammia2009,Menicucci2010}.  These all have the advantage of deterministic preparation over their optical qubit counterparts~\cite{Nielsen2004,Browne2005,Duan2005}, which rely on nondeterministic interactions and postselection to generate the desired state.  Each has specific advantages.

The \emph{canonical method}~\cite{Menicucci2006} involves single-mode squeezers and controlled-$Z$~($\CZ$) gates, which are an example of a quantum nondemolition~(QND) interaction~\cite{Bachor2004}.  The $\CZ$~gate can be implemented using beamsplitters and inline squeezing (i.e.,~squeezing of a state other than the vacuum)~\cite{Braunstein2005,Yurke1985}, which is experimentally challenging but achievable using current technology~\cite{Yoshikawa2008}.  All $\CZ$~gates commute, and this gate is the natural CV generalization of its qubit counterpart (used to create the qubit cluster states~\cite{Briegel2001}), thus making this design particularly amenable to theoretical analysis.  

The \emph{linear-optics method}~\cite{vanLoock2007} eliminates the need for inline squeezing by replacing the $\CZ$~gates with a network of beamsplitters.  The replacement is not one for one, however, and an entirely new network will generally be needed to make a cluster state with a different graph (even if they differ by just one node).  The major advantage of this method is that only vacuum (i.e.,~offline) squeezing and linear optics are needed, thus making this the method of choice in experiments to date~\cite{Su2007,Yonezawa2010,Yukawa2008,Ukai2010,Miwa2010}.

The \emph{single-OPO method}~\cite{Menicucci2008,Flammia2009} combines all squeezing and interferometry into a single optical parametric oscillator~(OPO), encoding the entire cluster state within a single beam.  Each mode is an individual frequency within an optical frequency comb.  The advantage of this method is its scalability.  While the initial implementation is more complex than the linear-optics method, once the technology is established~\cite{Pooser2005,Pysher2010,Pysher2009a,Midgley2010}, it is in principle much easier to scale up to cluster states that are larger by several orders of magnitude~\cite{Flammia2009}.

The \emph{single-QND-gate method}~\cite{Menicucci2010} reintroduces the experimentally challenging $\CZ$~gate.  But in this case, only \emph{one} such gate is needed because the modes are encoded \emph{temporally}, each traversing the same optical path (but at different times) and each passing multiple times through the same optical hardware implementing the $\CZ$~gate.  This method has the additional advantage that the cluster state is extended as needed---simultaneously with measurements implementing an algorithm on it---in a manner analogous to repeatedly laying down additional track in front of a moving train car (a ``Wallace and Gromit'' approach; see footnote in Ref.~\cite{Menicucci2010}).  Such a method eliminates the need for long-time coherence of a large cluster state because only a small piece of the state exists at any given time.

The current proposal is based on key elements from the last three.  The linear-optics method~\cite{vanLoock2007} is used to generate states produced by the single-OPO method~\cite{Menicucci2008,Flammia2009} using the temporal-mode encoding of the single-QND-gate method~\cite{Menicucci2010}.  The current method improves over the single-QND-gate method by eliminating the need for inline squeezing (in the $\CZ$~gate). The key observation that makes this simplification possible is that the states produced in the single-OPO method are Gaussian projected entangled pair states (GPEPS)~\cite{Ohliger2010}.  GPEPS states can be described as a collection of entangled pairs linked together by projecting adjacent ends down to a lower-dimensional joint subspace (schematically,~%
\begin{tikzpicture} [baseline=-0.5*\micronodesize,x=5mm, y=5mm, inner sep=0pt,outer sep=0pt]
	\matrix [column sep=0.5*\micronodesize] {
		\path node (a1) [micro] {} ++(1,0) node (b1) [micro] {}; \draw [poslink] (a1) -- (b1); &
		\path node (a2) [micro] {} ++(1,0) node (b2) [micro] {}; \draw [poslink] (a2) -- (b2); \\
	};
	\node (link) [ellipse,draw=red,fit=(a2) (b1)] {};
\end{tikzpicture}
becomes~%
\begin{tikzpicture} [baseline=-0.5*\micronodesize,x=5mm, y=5mm, inner sep=0pt,outer sep=0pt]
	\path node (a) [micro] {} ++(1,0) node (b) [micro] {} ++(1,0) node (c) [micro] {};
	\draw [poslink] (a) -- (b) -- (c);
\end{tikzpicture}
after projection on the circled nodes).

The paper is organized as follows.  In Section~\ref{sec:graphs}, I introduce the graphical formalism used for the main derivations.  Section~\ref{sec:GPEPSqw} describes a temporal-mode GPEPS-based construction of a CV quantum wire, and Section~\ref{sec:GPEPSlattice} follows with the same for a square-lattice CV cluster state.  Section~\ref{sec:discussion} concludes with some discussion, including comments on fault tolerance and error correction in the CV one-way QC model.

\section{Graphs for Gaussian pure states}
\label{sec:graphs}

\subsection{Basic properties}
\label{subsec:graphs:props}

To describe the states used in this work, I will use the graphical formalism that covers all Gaussian pure states in a unique and unified fashion~\cite{Menicucci2010b}.  This formalism is summarized here and simplified for the current presentation.  The usual graphical formalism for CV cluster states~\cite{Gu2009,Zhang2008a,Zhang2010} uses graphs with real-valued weights and is limited to representing ideal CV cluster states, which are necessarily infinitely squeezed and thus unphysical.  Using the complex-matrix representation of Gaussian states~\cite{Simon1988}, the graphical calculus for Gaussian pure states~\cite{Menicucci2010b} generalizes these ideal CV cluster-state graphs to complex-weighted graphs that can be used to uniquely represent \emph{any} Gaussian pure state,\footnote{The formalism only describes the noise properties of Gaussian pure states~\cite{Menicucci2010b}.  Thus, overall displacements are not represented.  From now on, I will assume that this caveat is understood.} including approximate CV cluster states.  In addition, other types of CV graphs, such as the $\cH$-graphs used in the single-OPO method~\cite{Menicucci2007,Flammia2009}, are incorporated directly into the formalism.  Local and two-local Gaussian unitary operations, as well as local quadrature measurements, can be visualized as graph transformations.

To every $N$-mode Gaussian pure state~$\ket{\psi_{\mat Z}}$ (ignoring overall displacement and up to an overall phase) we can uniquely assign an undirected, complex-wieghted graph on $N$~nodes whose adjacency matrix~$\mat Z$ is an~$N\times N$ complex symmetric matrix with positive-definite imaginary part.  That is,
\begin{align}
\label{eq:}
	\mat Z = i\mat U + \mat V\,,
\end{align}
where~$\mat U = \mat U^\tp$ and~$\mat V = \mat V^\tp$ are real, and~$\mat U > 0$.  All Gaussian pure states have a uniquely associated graph of this form, and all graphs satisfying these conditions are uniquely associated with a Gaussian pure state~\cite{Menicucci2010b}.  This matrix arises naturally (up to an overall factor of~$i$) in the position-space wavefunction for the state:
\begin{align}
\label{eq:Gausswavefunction}
	\psi_{\mat Z}(\vec q) &= \frac {(\det \mat U)^{1/4}} {\pi^{N/4}} \exp \left[- \frac 1 2 \vec q^\tp (\mat U - i\mat V) \vec q \right]\,,
\end{align}
where $\vec q$~is a column vector of c-numbers.  Henceforth, I will unambiguously refer to ``the Gaussian graph~$\mat Z$.''

All Gaussian unitary operations can be represented by the action of a symplectic matrix
\begin{align}
	\mat S =
	\begin{pmatrix}
		\mat A	& \mat B	\\
		\mat C	& \mat D
	\end{pmatrix}
\end{align}
on a column vector of Heisenberg-picture operators~$\opvec x = \left( \begin{smallmatrix} \opvec q \\ \opvec p \end{smallmatrix} \right)$, where~$\opvec q = (\op q_1, \dotsc, \op q_N)^\tp$, and~$\opvec p = (\op p_1, \dotsc, \op p_N)^\tp$, with $\opvec x' = \mat S \mat x$.  The new Gaussian graph~$\mat Z'$ associated to the resulting Gaussian pure state satisfies
\begin{align}
\label{eq:SonZ}
	\mat Z \xmapsto{\quad\mat S\quad} \mat Z' = (\mat C + \mat D \mat Z) (\mat A + \mat B \mat Z)^{-1}\,,
\end{align}
which can be interpreted as a (generally rather complicated) graph transformation rule.  However, since all general Gaussian unitary operations can be decomposed into a sequence of local and 2-local Gaussian unitaries chosen from a fiducial set, we can build up complicated graph transformations by repeated application of simpler ones.  Ref.~\cite{Menicucci2010b} has more details on the general case; we will use a simplified version.

The Gaussian pure state~$\ket{\psi_{\mat Z}}$ satisfies the following \emph{nullifier relation} with respect to its graph~$\mat Z$:
\begin{align}
\label{eq:nullifier}
	(\opvec p - \mat Z \opvec q) \ket{\psi_{\mat Z}} = 0\,,
\end{align}
where the entries in~$\opvec p$ and~$\opvec q$ are to be interpreted as Schr\"odinger-picture operators in this context because the state~$\ket{\psi_{\mat Z}}$ is indicated explicitly.\footnote{Notice that this is a literal equality, unlike the relation of similar form, $\opvec p - \mat A \opvec q \to 0$, for ideal CV cluster states with graph~$\mat A$, which is strictly true only in the limit of infinite squeezing~\cite{Gu2009}.}  The vacuum (i.e.,~ground state) on all~$N$ modes is represented by~$\mat Z = i\mat \id$, a completely disconnected graph with only self-loops of weight~$i$.  It is clear that Eq.~\eqref{eq:nullifier} holds for this state because~$\op p_j - i \op q_j = -i\sqrt 2 \op a_j$, where $\op a_j$~is the annihilation operator for mode~$j$.  In the limit that $\mat U \to \mat 0$, the Gaussian graph~$\mat Z$ limits to a real-weighted ideal CV cluster-state graph~$\mat V$, and Eq.~\eqref{eq:nullifier} becomes the usual nullifier relation for these states~\cite{Gu2009}.  For our purposes, we would also like to represent the $\cH$-graph states of the single-OPO method~\cite{Menicucci2007}.  These states have~$\mat V = \mat 0$.  Given an~$\cH$-graph~$\mat G$, the associated Gaussian graph is
\begin{align}
\label{eq:ZHgraph}
	\mat Z = ie^{-2\alpha \mat G}\,,
\end{align}
where $\alpha > 0$~is an overall squeezing parameter.

$\cH$-graphs are incorporated into this graphical formalism through the matrix exponential map, as in Eq.~\eqref{eq:ZHgraph}, which is normally a highly nontrivial operation.  However, it simplifies under certain conditions.  First of all, note that when $\mat G$~is \emph{bipartite} (i.e.,~its nodes can be separated completely into two groups such that no edge links nodes of the same group), it can be written as
\begin{align}
\label{eq:Gselfinv}
	\mat G = 
	\begin{pmatrix}
		\mat 0	& \mat G_0^\tp	\\
		\mat G_0	& \mat 0
	\end{pmatrix}
	\,,
\end{align}
When $\mat G$~is additionally \emph{self-inverse} (i.e.,~$\mat G^2 = \mat I$), then $\mat G_0$~is square and satisfies~$\mat G_0^\tp \mat G_0 = \mat G_0 \mat G_0^\tp = \mat \id$.  In this case, Eq.~\eqref{eq:ZHgraph} simplifies to
\begin{align}
\label{eq:ZfromselfinvG}
	\mat Z &= i \cosh 2\alpha\, \mat \id - i \sinh 2\alpha\, \mat G \nonumber \\
	&=
	\begin{pmatrix}
		i\cosh 2\alpha\, \mat \id	& -i\sinh 2\alpha\, \mat G_0^\tp	\\
		-i\sinh 2\alpha\, \mat G_0	& i\cosh 2\alpha\, \mat \id
	\end{pmatrix}
	\,.
\end{align}
The essence of the $\cH$-graph construction method is that this state is equivalent to an approximate CV cluster state after phase shifting one of the two sets of nodes (with respect to the graph bipartition)~\cite{Menicucci2007,Flammia2009}.  Performing a Fourier transform (i.e.,~phase shift by~$-\tfrac \pi 2$) on the first half of the nodes gives~\cite{Menicucci2010b}
\begin{align}
\label{eq:CVCSfromG}
	\mat Z'
	&=
	\begin{pmatrix}
		i\sech 2\alpha\, \mat \id	& \tanh 2\alpha\, \mat G_0^\tp	\\
		\tanh 2\alpha\, \mat G_0	& i\sech 2\alpha\, \mat \id
	\end{pmatrix}
	\,,
\end{align}
which is indeed an approximate CV cluster state since~$\mat Z' \to \mat G$ in the infinite-squezing limit ($\alpha \to \infty$).\footnote{Note that this is an approximate CV cluster state despite the fact that it is \emph{inequivalent} to the analogous approximate CV cluster state made by the canonical method~\cite{Menicucci2006}, whose Gaussian graph would be~$\mat Z = ie^{-2r}\mat \id + \mat G$.  In this expression, $r$~is the single-mode squeezing parameter for all the nodes that then pass through a collection of $\CZ$~gates in accord with~$\mat G$.  One major advantage of the unified Gaussian graphical formalism is the ability to distinguish between these distinct approximants to the same ideal CV cluster state.}  The fact that the ideal CV cluster state approximated by~$\mat Z'$ has the same graph as the $\cH$-graph~$\mat G$ is a peculiar feature of bipartite, self-inverse~$\cH$-graphs~\cite{Zaidi2008,Flammia2009}; usually the graphs are very different~\cite{Menicucci2007}.

\subsection{\texorpdfstring{$\cH$}{H}-graph states made without an \texorpdfstring{$\cH$}{H}-graph}
\label{subsec:graphs:noHgraph}

At this point, we should forget about the $\cH$-graph method of construction for~$\mat Z$.  It doesn't matter how the state represented by Eq.~\eqref{eq:ZfromselfinvG} was created---all we care about is that the graph for it is given by~$\mat Z$ as shown, and we will use graphs of this form even when $\mat G$~is only \emph{approximately} self-inverse (due to imperfections at the boundary of a line or lattice).  This is an important distinction because the simplification of the exponential map only holds when $\mat G$ is \emph{exactly} self-inverse.  Stray connections and imperfections in the $\cH$-graph will generally ``bleed out'' via the exponential and contaminate connections all over the Gaussian graph~$\mat Z$, which will end up failing to produce the state that we want, Eq.~\eqref{eq:ZfromselfinvG}, and without an easy way to isolate and eliminate the bad links.  If instead we start directly from Eq.~\eqref{eq:ZfromselfinvG}, then we'll find---when we phase shift the nodes as above---that imperfections in~$\mat Z$ are confined to a small neighborhood of the imperfections in~$\mat G$.  In fact, in this case, we have~\cite{Menicucci2010b}
\begin{align}
\label{eq:CVCSfromGimperfect}
	&\mat Z' = \nonumber \\
	&
	\begin{pmatrix}
		i\sech 2\alpha\, \mat \id	& \tanh 2\alpha\, \mat G_0^\tp	\\
		\tanh 2\alpha\, \mat G_0	& i\sech 2\alpha [\cosh^2 2\alpha\, \mat \id - \sinh^2 2\alpha\, \mat G_0 \mat G_0^\tp]
	\end{pmatrix}
	\,,
\end{align}
which reduces to Eq.~\eqref{eq:CVCSfromG} when $\mat G$~is self-inverse.

Consider the case in which only a small collection of nodes prevents~$\mat G$ from being strictly self-inverse.  Because the matrix~$\mat Z'$ in Eq.~\eqref{eq:CVCSfromGimperfect} involves the \emph{square} of~$\mat G$ (through the term~$\mat G_0 \mat G_0^\tp$), nodes with incorrect edges in~$\mat G$ can only affect other nodes that are at most two steps away (rather than everywhere, in the case of the exponential).  To see this, define~$\mat G = \bar{\mat G} + \mat E$, where $\bar{\mat G}$~is bipartite and exactly self-inverse, and $\mat E$~is bipartite (with the same partitioning), real, symmetric, and sparse.  In addition, both~$\bar{\mat G}$ and~$\mat E$ have an analogous decomposition to that of~$\mat G$ in Eq.~\eqref{eq:Gselfinv}, and therefore~$\mat G_0 = \bar{\mat G}_0 + \mat E_0$, as well.   $\mat E$~is a matrix of just a few ``errors'' that cause~$\mat G$ to fail to be perfectly self-inverse.  We have the following relation:
\begin{align}
\label{eq:G0G0}
	\mat G_0 \mat G_0^\tp &= (\bar{\mat G}_0 + \mat E_0) (\bar{\mat G}_0 + \mat E_0)^\tp \nonumber \\
	&= \mat \id + \mat E_0 \bar{\mat G}_0^\tp + \bar{\mat G}_0 \mat E_0^\tp + \mat E_0 \mat E_0^\tp\,.
\end{align}
Now define~$\mat P_0$ to be a matrix with the same dimensions as~$\mat E_0$ that is constructed in the following way: first, place a~1 in the diagonal entry~$(P_0)_{jj}$ if row~$j$ of~$\mat E_0$ consists of all zeros (i.e.,~if $(E_0)_{jk} = 0\; \forall k$), and place 0's everywhere else; then, remove the all-zero rows of~$\mat P_0$.  Notice that~$\mat P_0 \mat E_0 = \mat 0$, which gives
\begin{align}
\label{eq:P0onG0}
	\mat P_0 \mat G_0 \mat G_0^\tp \mat P_0^\tp = \mat P_0 \mat P_0^\tp = \mat \id\,,
\end{align}
where the identity matrix on the right is smaller than that in Eq.~\eqref{eq:G0G0}.  Now further define
\begin{align}
	\mat P :=
	\begin{pmatrix}
		\mat \id	& \mat 0	\\
		\mat 0	& \mat P_0
	\end{pmatrix}
	\,,
\end{align}
which is also not square, where the identity matrix is the same size as~$\mat E_0$, and the blocks of~0 are sized appropriately.\footnote{Alternatively, $\mat P$~can be defined directly in terms of the Moore-Penrose pseudoinverse of~$\mat E_0$, which is denoted~$\mat E_0^+$.  Starting with
\begin{align*}
	\begin{pmatrix}
		\mat \id	& \mat 0	\\
		\mat 0	& \mat \id - \mat E_0 \mat E_0^+
	\end{pmatrix}
	\,,
\end{align*}
which \emph{is} square, we can form~$\mat P$ simply by removing the all-zero rows.}

Measurements of~$\op q_j$ delete a node~$j$ and all of its links from a given Gaussian graph~\cite{Menicucci2010b}, which is equivalent to deleting the $j^\text{th}$~row and column from its adjacency matrix.  Conjugating $\mat Z'$~from Eq.~\eqref{eq:CVCSfromGimperfect} by~$\mat P$ deletes from~$\mat Z'$ the rows and columns that correspond to the nonzero rows of~$\mat E_0$ (and, equivalently, the nonzero columns of~$\mat E_0^\tp$):
\begin{align}
	\mat P \mat Z' \mat P^\tp &=
	\begin{pmatrix}
		i\sech 2\alpha\, \mat \id			& \tanh 2\alpha\, \mat G_0^\tp \mat P_0^\tp	\\
		\tanh 2\alpha\, \mat P_0 \mat G_0	& i\sech 2\alpha\, \mat \id
	\end{pmatrix}
	\nonumber \\
	&= i\sech 2\alpha\, \mat \id + \tanh 2\alpha\, \mat P \mat G \mat P^\tp
	\,,
\end{align}
where Eq.~\eqref{eq:P0onG0} has been used.  This is an approximate CV cluster state with most (but not all) of the graph~$\mat G$ intact; the troublesome links have been deleted by deleting nodes that they attach to.  In this case, all of the deleted nodes are taken from the second set of the graph bipartition.   When these nodes are located near the boundary of a regular lattice (or otherwise isolated in a graph with local topology), then deleting these nodes (or ones just to the ``inside'' of them) using $\op q$~measurements is all that is needed to eliminate the imperfections.

The point of the preceding analysis is to justify ignoring the irregular boundary conditions in the sections that follow.  In the single-OPO scheme, these boundary irregularities in~$\mat G$ could not be ignored because they would contaminate the entire state through the exponential map.  In this case, they can be dealt with simply by deleting nearby nodes.  This is analogous to using scissors to clean up the jagged edge of a torn piece of paper.

\subsection{Simplified graphical formalism}
\label{subsec:graphs:rules}

There are only three graph transformation rules that we will need for this work: (1)~measurements of~$\op q$, (2)~phase-plane inversion (i.e.,~phase shift by~$\pi$), and (3)~interference through a 50:50 beamsplitter.  All of the graphs we use will be of the form of Eq.~\eqref{eq:ZfromselfinvG}, but we will make the following modifications to the general formalism~\cite{Menicucci2010b}.  First, we will not draw the self-loops indicated by the diagonal of~$\mat Z$, but we will be mindful of them when we derive the simplified graph rules that follow.  Second, colors indicate the sign of the edge weight but not its actual value, with
\begin{align}
\label{eq:Zweights}
	\tikz {\draw [yshift=2pt,poslink] (0,0) -- (4mm,0); \draw [yshift=-2pt,neglink] (0,0) -- (4mm,0);} = \mp i \cC \sinh 2\alpha \qquad \text{in the graph~$\mat Z$}\,,
\end{align}
where~$\cC>0$.  Nevertheless, for the graphs used in this work, the following relation often (but not always) holds:
\begin{align}
\label{eq:Cdef}
	\cC = \max(d_1,d_2)^{-\frac 1 2}\,,
\end{align}
where $d_{1,2}$~are, respectively, the degrees (number of neighbors) of the two nodes linked by the edge in question.%
\footnote{This seemingly strange formula arises because of the need for~$\mat G$ in Eq.~\eqref{eq:ZfromselfinvG} to be self-inverse almost everywhere (see Section~\ref{subsec:graphs:noHgraph}).  The exception to Eq.~\eqref{eq:Cdef} is when $\op q$~measurements are used to delete nodes and their edges from the graph; the remaining edges retain their weights before the deletion.}  %
Within the approximate CV cluster state~$\mat Z'$ that eventually results after the phase shifts, these weights become
\begin{align}
\label{eq:Zpweights}
	\tikz {\draw [yshift=2pt,poslink] (0,0) -- (4mm,0); \draw [yshift=-2pt,neglink] (0,0) -- (4mm,0);} &\mapsto \pm \cC \tanh 2\alpha \xrightarrow{\alpha \to \infty} \pm \cC \qquad \text{in the graph~$\mat Z'$} \,,
\end{align}
with $\alpha \to \infty$~indicating the infinite-squeezing limit.  The fact that only the sign of the weights is indicated visually does not take away from the rigor of these results; it is merely for visual simplicity.  When relevant, the actual edge weights will be specified.

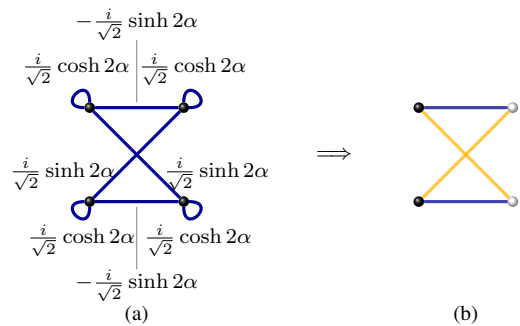
\begin{figure}[t!]
\begin{center}
\beginpgfgraphicnamed{graphics/graphsimple}%
\begin{tikzpicture} [scale=1,label distance=-4pt, pin distance=3.2em]

	\footnotesize

	\def\hgraph{(0,0)}
	\def\simplegraph{(3.5*\scale,0)}
	
	\def\labelloc{(-0.5*\scale,-2.2*\scale)}
	
	\def\symbolloc{(1.6*\scale,-0.5*\scale)}

	\def\scale{1.25}

	\path \hgraph +\labelloc node {(a)};
	\path \hgraph [outer sep=0pt]
		node [micro] (1) {}
		++(180:\scale) node [micro] (2) {}
		++(-90:\scale) node [micro] (3) {}
		++(0:\scale) node [micro] (4) {};

	\path (1)
		edge [fulllink,in=0,out=90, min distance=0.4cm] node [xshift=-2pt,above=0pt] {$\tfrac {i} {\sqrt 2}\cosh 2\alpha$} (1)
		edge [fulllink] node [inner sep=0, outer sep=0.5pt, above, pin={[outer sep=0,inner sep=0] 90:$-\tfrac {i} {\sqrt 2}\sinh 2\alpha$}] {} (2)
		edge [fulllink] node [xshift=-7pt,below left=-2pt] {$\tfrac {i} {\sqrt 2}\sinh 2\alpha$} (3);

	\path (2)
		edge [fulllink,in=90,out=180, min distance=0.4cm] node [above=0pt] {$\tfrac {i} {\sqrt 2}\cosh 2\alpha$} (2)
		edge [fulllink] node [xshift=9pt,below right=-2pt] {$\tfrac {i} {\sqrt 2}\sinh 2\alpha$} (4);
	
	\path (3)
		edge [fulllink,in=180,out=270, min distance=0.4cm] node [xshift=2pt, below=0pt] {$\tfrac {i} {\sqrt 2}\cosh 2\alpha$} (3)
		edge [fulllink] node [inner sep=0, outer sep=0.5pt, below, pin={[outer sep=0,inner sep=0] -90:$-\tfrac {i} {\sqrt 2}\sinh 2\alpha$}] {} (4);
	
	\path (4)
		edge [fulllink,in=270,out=0, min distance=0.4cm] node [below=0pt] {$\tfrac {i} {\sqrt 2}\cosh 2\alpha$} (4);

	\path \simplegraph +\labelloc node {(b)};
	\path \simplegraph [outer sep=0pt]
		node [micro-odd] (1) {}
		++(180:\scale) node [micro-even] (2) {}
		++(-90:\scale) node [micro-even] (3) {}
		++(0:\scale) node [micro-odd] (4) {};

	\path (1)
		edge [poslink] (2)
		edge [neglink] (3);

	\path (2)
		edge [neglink] (4);
	
	\path (3)
		edge [poslink] (4);

	\path \symbolloc node {$\Longrightarrow$};

\end{tikzpicture}%
\endpgfgraphicnamed%
\caption{\label{fig:graphsimple}Simplified graphical formalism for the Gaussian pure states represented in this paper.  (a)~Full graph for a particular Gaussian pure state~\cite{Menicucci2010b}.  (b)~Simplified graph for the same state, as used in this paper.  Self loops are not drawn; colored edges indicate the sign of the weights as in Eq.~\eqref{eq:Zweights}, with~$\cC = \tfrac 1 {\sqrt 2}$ (see text) for all edges since each edge connects nodes having exactly $2$~neighbors; and colored nodes indicate the bipartite splitting, with white nodes receiving a Fourier transform to convert this to an approximate CV cluster-state with the same graph, but now with weights given by Eq.~\eqref{eq:Zpweights}.}
\end{center}
\end{figure}

Finally, the bipartite nature of~$\mat Z$ will be indicated by coloring the nodes \text{\evencolor} and \text{\oddcolor}, according to the division into two sets.  This coloring is only a visual distinction and has no physical significance on its own.\footnote{In particular, while different colors were used in Ref.~\cite{Menicucci2010b} to distinguish the meaning of nodes in ideal CV cluster-state graphs from their meaning in general Gaussian graphs, both colors of nodes in the present paper correspond to black nodes in Ref.~\cite{Menicucci2010b}.  All of the graphs used here are simplified versions of general Gaussian graphs.}  The \text{\oddcolor} nodes will eventually be Fourier transformed (i.e.,~phase shifted by~$-\tfrac \pi 2$) to implement the map~$\mat Z \mapsto \mat Z'$, as in Eq.~\eqref{eq:CVCSfromG}, but since all of the graphs we will discuss satisfy~$\mat Z \sim \mat Z' \sim \mat G$ up to an affine transformation (and neglecting boundary irregularities), we will not need a graph transformation for the Fourier transform.  In addition, since CV cluster states are used for one-way QC by making local measurements only, the Fourier transform can be absorbed into the eventual measurement procedure and becomes simply a change of basis.\footnote{\label{foot:Fourier}Therefore, when we talk about, for example, measuring a \text{\oddcolor} node of the resulting CV cluster state in~$\op q$, we really mean measuring the same node of the original state in~$\op F^\dag \op q \op F = -\op p$.}  An example of this simplification of the Gaussian graph formalism is presented in Figure~\ref{fig:graphsimple}.

Below are the graph transformations used in this work.  These rules are rigorous for the graphs considered in this work and are derived from the general transformation rules for Gaussian graphs~\cite{Menicucci2010b}, but they have been adapted and simplified for the representation used here.  These rules function in exactly the same way on both \text{\evencolor} and \text{\oddcolor} nodes.

\textbf{Measurement of~$\op q$}---As discussed above, this operation simply deletes the measured node and its links from the Gaussian graph~$\mat Z$.  The measured node is indicated by a small arrow:
\begin{align*}
\beginpgfgraphicnamed{graphics/qmeas}%
\begin{tikzpicture} [scale=1,label distance=-4pt, pin distance=3.2em]
%
	\footnotesize
	\def\orig{(0,0)}
	\def\newgraph{(2*\scale,0)}
	\def\symbolloc{(0.5*\scale,-0.5*\scale)}
	\def\scale{1}
%
	\path \orig [outer sep=0pt]
		node [micro-odd] (1) {}
		++(180:\scale) node [micro-even] (2) {}
		++(-90:\scale) node [micro-even] (3) {}
		++(0:\scale) node [micro-odd] (4) {};
	\path (1)
		edge [poslink] (2)
		edge [neglink] (3);
	\path (2)
		edge [neglink] (4);
	\path (3)
		edge [poslink] (4);
	\path (2) [draw,<-]
		++(135:4pt) -- ++(135:1.5ex)
		;
	\path \symbolloc node {$\longmapsto$};
%
	\path \newgraph [outer sep=0pt]
		node [micro-odd] (1) {}
		++(180:\scale) node [micro-even, \phantomfade] (2) {}
		++(-90:\scale) node [micro-even] (3) {}
		++(0:\scale) node [micro-odd] (4) {};
	\path (1)
		edge [phantomlink] (2)
		edge [neglink] (3);
	\path (2)
		edge [phantomlink] (4);
	\path (3)
		edge [poslink] (4);
\end{tikzpicture}\,.%
\endpgfgraphicnamed%
\end{align*}
Recall that the colors only indicate the sign of the edge weights.  The actual values of these weights do not change with the deletion, but it is worth emphasizing that they may no longer satisfy Eq.~\eqref{eq:Cdef}.

\textbf{Phase-plane inversion (phase shift by~$\pi$)}---This operation is also very simple.  All edges attached to the inverted node (indicated by highlighting) change color:
\begin{align*}
\beginpgfgraphicnamed{graphics/inversion}%
\begin{tikzpicture} [scale=1,label distance=-4pt, pin distance=3.2em]
%
	\footnotesize
	\def\orig{(0,0)}
	\def\newgraph{(2*\scale,0)}
	\def\symbolloc{(0.5*\scale,-0.5*\scale)}
	\def\scale{1}
%
	\path \orig [outer sep=0pt]
		node [micro-odd] (1) {}
		++(180:\scale) node [micro-even] (2) {}
		++(-90:\scale) node [micro-even] (3) {}
		++(0:\scale) node [micro-odd] (4) {};
	\path (1)
		edge [poslink] (2)
		edge [neglink] (3);
	\path (2)
		edge [neglink] (4);
	\path (3)
		edge [poslink] (4);
	\draw [nodehighlight] (2) circle (\micronodesize);
	\path \symbolloc node {$\longmapsto$};
%
	\path \newgraph [outer sep=0pt]
		node [micro-odd] (1) {}
		++(180:\scale) node [micro-even] (2) {}
		++(-90:\scale) node [micro-even] (3) {}
		++(0:\scale) node [micro-odd] (4) {};
	\path (1)
		edge [neglink] (2)
		edge [neglink] (3);
	\path (2)
		edge [poslink] (4);
	\path (3)
		edge [poslink] (4);
\end{tikzpicture}\,.%
\endpgfgraphicnamed%
\end{align*}
This corresponds to the edge weight changing sign but not magnitude.

\textbf{50:50 beamsplitter}---The 50:50 beamsplitter implements~$\mat S_{\text{BS}}(\tfrac \pi 4)$ from Ref.~\cite{Menicucci2010b}:
\begin{align}
\label{eq:SBSpi4}
	\mat S_{\text{BS}}(\tfrac \pi 4) =
	\begin{pmatrix}
		\frac {1} {\sqrt 2}	&	-\frac {1} {\sqrt 2}	&	0		&	0		\\
		\frac {1} {\sqrt 2}	&	\frac {1} {\sqrt 2}	&	0		&	0		\\
		0		&	0		&	\frac {1} {\sqrt 2}	&	-\frac {1} {\sqrt 2}	\\
		0		&	0		&	\frac {1} {\sqrt 2}	&	\frac {1} {\sqrt 2}
	\end{pmatrix}
	\,.
\end{align}
Under the conditions of no link between the nodes being interfered and equal self-loop weights (which is the case with all graphs in this paper but is not displayed in the simplified formalism), we obtain the following very simple rules:
\begin{center}
\beginpgfgraphicnamed{graphics/beamsplit}%
\begin{tabular}{c@{\qquad\qquad}c}
\begin{tikzpicture} [scale=1,label distance=-4pt, pin distance=3.2em]

	\footnotesize

	\def\orig{(0,0)}
	\def\newgraph{(2*\scale,0)}
	
	\def\symbolloc{(0.5*\scale,0)}

	\def\scale{1}

	\path \orig [outer sep=0pt]
		node [micro-odd] (1) {}
		+(160:\scale) node [micro-even] (2) {}
		+(-160:\scale) node [micro-even] (3) {}
		;

	\path (1)
		edge [poslink] (2)
		;
	
	\draw [red, ->, shorten >=2pt, shorten <=2pt, bend right] (2) to (3);
		
	\path \symbolloc node {$\longmapsto$};

	\path \newgraph [outer sep=0pt]
		node [micro-odd] (1) {}
		+(160:\scale) node [micro-even] (2) {}
		+(-160:\scale) node [micro-even] (3) {}
		;

	\path (1)
		edge [poslink] (2)
		edge [poslink] (3)
		;

\end{tikzpicture}%
&
\begin{tikzpicture} [scale=1,label distance=-4pt, pin distance=3.2em]

	\footnotesize

	\def\orig{(0,0)}
	\def\newgraph{(2*\scale,0)}
	
	\def\symbolloc{(0.5*\scale,0)}

	\def\scale{1}

	\path \orig [outer sep=0pt]
		node [micro-odd] (1) {}
		+(160:\scale) node [micro-even] (2) {}
		+(-160:\scale) node [micro-even] (3) {}
		;

	\path (1)
		edge [neglink] (2)
		;
	
	\draw [red, ->, shorten >=2pt, shorten <=2pt, bend right] (2) to (3);
		
	\path \symbolloc node {$\longmapsto$};

	\path \newgraph [outer sep=0pt]
		node [micro-odd] (1) {}
		+(160:\scale) node [micro-even] (2) {}
		+(-160:\scale) node [micro-even] (3) {}
		;

	\path (1)
		edge [neglink] (2)
		edge [neglink] (3)
		;

\end{tikzpicture}\,\phantom{.}%
\\
\\
\begin{tikzpicture} [scale=1,label distance=-4pt, pin distance=3.2em]

	\footnotesize

	\def\orig{(0,0)}
	\def\newgraph{(2*\scale,0)}
	
	\def\symbolloc{(0.5*\scale,0)}

	\def\scale{1}

	\path \orig [outer sep=0pt]
		node [micro-odd] (1) {}
		+(160:\scale) node [micro-even] (2) {}
		+(-160:\scale) node [micro-even] (3) {}
		;

	\path (1)
		edge [poslink] (3)
		;
	
	\draw [red, ->, shorten >=2pt, shorten <=2pt, bend right] (2) to (3);
		
	\path \symbolloc node {$\longmapsto$};

	\path \newgraph [outer sep=0pt]
		node [micro-odd] (1) {}
		+(160:\scale) node [micro-even] (2) {}
		+(-160:\scale) node [micro-even] (3) {}
		;

	\path (1)
		edge [neglink] (2)
		edge [poslink] (3)
		;

\end{tikzpicture}%
&
\begin{tikzpicture} [scale=1,label distance=-4pt, pin distance=3.2em]

	\footnotesize

	\def\orig{(0,0)}
	\def\newgraph{(2*\scale,0)}
	
	\def\symbolloc{(0.5*\scale,0)}

	\def\scale{1}

	\path \orig [outer sep=0pt]
		node [micro-odd] (1) {}
		+(160:\scale) node [micro-even] (2) {}
		+(-160:\scale) node [micro-even] (3) {}
		;

	\path (1)
		edge [neglink] (3)
		;
	
	\draw [red, ->, shorten >=2pt, shorten <=2pt, bend right] (2) to (3);
		
	\path \symbolloc node {$\longmapsto$};

	\path \newgraph [outer sep=0pt]
		node [micro-odd] (1) {}
		+(160:\scale) node [micro-even] (2) {}
		+(-160:\scale) node [micro-even] (3) {}
		;

	\path (1)
		edge [poslink] (2)
		edge [neglink] (3)
		;

\end{tikzpicture}\,.%
\end{tabular}
\endpgfgraphicnamed%
\end{center}
Notice that the beamsplitter duplicates a single link, but multiplies the magnitude of each by a factor of~$\tfrac {1} {\sqrt 2}$ in the process.  The arrow points from node~1 to node~2 of the interaction, distinguishing the effect of the negative signs in Eq.~\eqref{eq:SBSpi4}.  When an edge is copied in the direction of the arrow, the same color is applied to the new edge.  When it is copied in the opposite direction, the new edge has the opposite color.  Even with their simplified presentation, all of these graph transformation rules are rigorous for the graphs used in this paper.

\section{GPEPS quantum wire}
\label{sec:GPEPSqw}

The single-OPO ``quantum wire''---a cluster state with a one-dimensional topology---actually has a width of two nodes and looks like this~\cite{Flammia2009}:
\begin{align}
\mat Z_{\text{OPO}} &=
\beginpgfgraphicnamed{graphics/OPOqw}%
\begin{tikzpicture} [baseline=-2.5mm-0.5ex, x=1cm, y=5mm, inner sep=0pt,outer sep=0pt]
	\def\lastn{5}
	\begin{scope}
	\clip ($ (.4,-1) - 0.55*(0,\micronodesize) $) rectangle ($ (\lastn+.6,0) + 1.1*(0,\micronodesize) $);
	\foreach \n in {0,...,\lastn}
	{
		\ifthenelse {\not\isodd{\n}}
			{
			\node (ia) [micro-even] at (\n,0) {};
			\node (ib) [micro-even] at (\n,-1) {};
			}
			{
			\node (ia) [micro-odd] at (\n,0) {};
			\node (ib) [micro-odd] at (\n,-1) {};
			}
		\node [micro-no-color] (oa) at (\n+1,0) {};
		\node [micro-no-color] (ob) at (\n+1,-1) {};
		\ifthenelse {\not\isodd{\n}}
			{\piminus {ia} {ib} {oa} {ob}}
			{\piplus {ia} {ib} {oa} {ob}
			;}
	}
		\ifthenelse {\isodd{\lastn}}
			{
			\node (ia) [micro-even] at (\lastn+1,0) {};
			\node (ib) [micro-even] at (\lastn+1,-1) {};
			}
			{
			\node (ia) [micro-odd] at (\lastn+1,0) {};
			\node (ib) [micro-odd] at (\lastn+1,-1) {};
			}
	\end{scope}
\end{tikzpicture}
\endpgfgraphicnamed%
\,[\cC =\tfrac 1 2] \,.
\end{align}
Notice that only a portion of the full graph is displayed.\footnote{If the full graph has periodic boundary conditions (or if it is formally infinite), then $\mat G$~for it is self-inverse~\cite{Flammia2009}, and Eq.~\eqref{eq:CVCSfromG} is satisfied.  If, however, the graph has at least one boundary, then only Eq.~\eqref{eq:CVCSfromGimperfect} is satisfied.  As we have already seen in Section~\ref{subsec:graphs:noHgraph}, these boundary effects can be dealt with by suitably clipping the graph using $\op q$~measurements.}  We can perform a $\pi$~phase shift on every other node on the top row to put this state into a translationally invariant form (ignoring node color, which has no physical significance):
\begin{align}
\mat Z &=
\beginpgfgraphicnamed{graphics/OPOqwhighlight}%
\begin{tikzpicture} [baseline=-2.5mm-0.5ex, x=1cm, y=5mm, inner sep=0pt,outer sep=0pt]
	\def\lastn{5}
	\begin{scope}
	\clip ($ (.4,-1) - 0.55*(0,\micronodesize) $) rectangle ($ (\lastn+.6,0) + 1.1*(0,\micronodesize) $);
	\foreach \n in {0,...,\lastn}
	{
		\ifthenelse {\not\isodd{\n}}
			{
			\node (ia) [micro-even] at (\n,0) {};
			\node (ib) [micro-even] at (\n,-1) {};
			}
			{
			\node (ia) [micro-odd] at (\n,0) {};
			\node (ib) [micro-odd] at (\n,-1) {};
			}
		\node [micro-no-color] (oa) at (\n+1,0) {};
		\node [micro-no-color] (ob) at (\n+1,-1) {};
		\ifthenelse {\not\isodd{\n}}
			{\piminus {ia} {ib} {oa} {ob}}
			{\piplus {ia} {ib} {oa} {ob}
			\draw [nodehighlight] (ia) circle (\micronodesize);}
	}
		\ifthenelse {\isodd{\lastn}}
			{
			\node (ia) [micro-even] at (\lastn+1,0) {};
			\node (ib) [micro-even] at (\lastn+1,-1) {};
			}
			{
			\node (ia) [micro-odd] at (\lastn+1,0) {};
			\node (ib) [micro-odd] at (\lastn+1,-1) {};
			}
	\end{scope}
\end{tikzpicture}
\endpgfgraphicnamed%
\,[\cC =\tfrac 1 2]
\nonumber \\
&=
\beginpgfgraphicnamed{graphics/qwtrans}%
\begin{tikzpicture} [baseline=-2.5mm-0.5ex, x=1cm, y=5mm, inner sep=0pt, outer sep=0pt]
	\def\lastn{5}
	\begin{scope}
	\clip ($ (.4,-1) - 0.55*(0,\micronodesize) $) rectangle ($ (\lastn+.6,0) + 0.55*(0,\micronodesize) $);
	\foreach \n in {0,...,\lastn}
	{
		\ifthenelse {\not\isodd{\n}}
			{
			\node (ia) [micro-even] at (\n,0) {};
			\node (ib) [micro-even] at (\n,-1) {};
			}
			{
			\node (ia) [micro-odd] at (\n,0) {};
			\node (ib) [micro-odd] at (\n,-1) {};
			}
		\node (oa) at (\n+1,0) {};
		\node (ob) at (\n+1,-1) {};
		\beamsplit {ia} {ib} {oa} {ob}
	}
		\ifthenelse {\isodd{\lastn}}
			{
			\node (ia) [micro-even] at (\lastn+1,0) {};
			\node (ib) [micro-even] at (\lastn+1,-1) {};
			}
			{
			\node (ia) [micro-odd] at (\lastn+1,0) {};
			\node (ib) [micro-odd] at (\lastn+1,-1) {};
			}
	\end{scope}
\end{tikzpicture}
\endpgfgraphicnamed%
\,[\cC =\tfrac 1 2] \,.
\end{align}
The connection to a GPEPS construction is immediate when we notice that this is exactly the result of applying the 50:50 beamsplitter transformation rules to a collection of two-mode squeezed states arranged as follows:
\begin{align}
\label{eq:TMSqw}
\mat Z &=
\beginpgfgraphicnamed{graphics/TMSqw}%
\begin{tikzpicture} [baseline=-2.5mm-0.5ex, x=1cm, y=5mm, inner sep=0pt, outer sep=0pt]
	\def\lastn{5}
	\begin{scope}
	\clip ($ (.4,-1) - 0.55*(0,\micronodesize) $) rectangle ($ (\lastn+.6,0) + 0.55*(0,\micronodesize) $);
	\foreach \n in {0,...,\lastn}
	{
		\ifthenelse {\not\isodd{\n}}
			{
			\node (ia) [micro-even] at (\n,0) {};
			\node (ib) [micro-even] at (\n,-1) {};
			}
			{
			\node (ia) [micro-odd] at (\n,0) {};
			\node (ib) [micro-odd] at (\n,-1) {};
			}
		\node (oa) at (\n+1,0) {};
		\node (ob) at (\n+1,-1) {};
		\tmslink {ia} {ib} {oa} {ob}
		\draw [red, ->, shorten >=1pt, shorten <=1pt, bend right] (ia) to (ib);
	}
	\end{scope}
\end{tikzpicture}
\endpgfgraphicnamed%
\,[\cC =1] \\
\label{eq:GPEPSqw}
&=
\beginpgfgraphicnamed{graphics/GPEPSqw}%
\begin{tikzpicture} [baseline=-2.5mm-0.5ex, x=1cm, y=5mm, inner sep=0pt, outer sep=0pt]
	\def\lastn{5}
	\begin{scope}
	\clip ($ (.4,-1) - 0.55*(0,\micronodesize) $) rectangle ($ (\lastn+.6,0) + 0.55*(0,\micronodesize) $);
	\foreach \n in {0,...,\lastn}
	{
		\ifthenelse {\not\isodd{\n}}
			{
			\node (ia) [micro-even] at (\n,0) {};
			\node (ib) [micro-even] at (\n,-1) {};
			}
			{
			\node (ia) [micro-odd] at (\n,0) {};
			\node (ib) [micro-odd] at (\n,-1) {};
			}
		\node (oa) at (\n+1,0) {};
		\node (ob) at (\n+1,-1) {};
		\beamsplit {ia} {ib} {oa} {ob}
	}
		\ifthenelse {\isodd{\lastn}}
			{
			\node (ia) [micro-even] at (\lastn+1,0) {};
			\node (ib) [micro-even] at (\lastn+1,-1) {};
			}
			{
			\node (ia) [micro-odd] at (\lastn+1,0) {};
			\node (ib) [micro-odd] at (\lastn+1,-1) {};
			}
	\end{scope}
\end{tikzpicture}
\endpgfgraphicnamed%
\,[\cC =\tfrac 1 2] \,.
\end{align}
Notice that the order of application of these beamsplitters does not matter; the rules commute and result in the same graph.  The Gaussian graph~$\mat Z$ is the desired output of a temporal-mode linear-optics construction method.  Once in this form, measurements of~$\op q$ can be be made on all upper nodes (following the implicit phase shift by~$-\frac \pi 2$ on all \text{\oddcolor} nodes; see Section~\ref{subsec:graphs:rules}), projecting the bottom row of nodes into an ordinary (one-dimensional) approximate CV quantum wire~\cite{Gu2009}:
\begin{align}
\label{eq:qwreduced}
\mat Z_{\text{qw}} &=
\beginpgfgraphicnamed{graphics/qwmeasuring}%
\begin{tikzpicture} [baseline=-2.5mm-0.5ex, x=1cm, y=5mm, inner sep=0pt, outer sep=0pt]
	\def\lastn{5}
	\begin{scope}
	\clip (.4,-1.5) rectangle (\lastn+.6,0.5);
	\foreach \n in {0,...,\lastn}
	{
		\ifthenelse {\not\isodd{\n}}
			{
			\node (ia) [micro-even] at (\n,0) {};
			\node (ib) [micro-even] at (\n,-1) {};
			}
			{
			\node (ia) [micro-odd] at (\n,0) {};
			\node (ib) [micro-odd] at (\n,-1) {};
			}
		\node (oa) at (\n+1,0) {};
		\node (ob) at (\n+1,-1) {};
		\beamsplit {ia} {ib} {oa} {ob}
		\draw [<-] (ia) ++(135:4pt) -- ++(135:1.5ex);
	}
		\ifthenelse {\isodd{\lastn}}
			{
			\node (ia) [micro-even] at (\lastn+1,0) {};
			\node (ib) [micro-even] at (\lastn+1,-1) {};
			}
			{
			\node (ia) [micro-odd] at (\lastn+1,0) {};
			\node (ib) [micro-odd] at (\lastn+1,-1) {};
			}
	\end{scope}
\end{tikzpicture}
\endpgfgraphicnamed%
\,[\cC =\tfrac 1 2] \nonumber \\
&=
\beginpgfgraphicnamed{graphics/qwreduced}%
\begin{tikzpicture} [baseline=-2.5mm-0.5ex, x=1cm, y=5mm, inner sep=0pt, outer sep=0pt]
	\def\lastn{5}
	\begin{scope}
	\clip ($ (.4,-1) - 0.55*(0,\micronodesize) $) rectangle ($ (\lastn+.6,0) + 0.55*(0,\micronodesize) $);
	\foreach \n in {0,...,\lastn}
	{
		\ifthenelse {\not\isodd{\n}}
			{
			\node (ia) [micro-even, \phantomfade] at (\n,0) {};
			\node (ib) [micro-even] at (\n,-1) {};
			}
			{
			\node (ia) [micro-odd, \phantomfade] at (\n,0) {};
			\node (ib) [micro-odd] at (\n,-1) {};
			}
		\node (oa) at (\n+1,0) {};
		\node (ob) at (\n+1,-1) {};
		\linearlink {ia} {ib} {oa} {ob}
	}
		\ifthenelse {\isodd{\lastn}}
			{
			\node (ia) [micro-even, \phantomfade] at (\lastn+1,0) {};
			\node (ib) [micro-even] at (\lastn+1,-1) {};
			}
			{
			\node (ia) [micro-odd, \phantomfade] at (\lastn+1,0) {};
			\node (ib) [micro-odd] at (\lastn+1,-1) {};
			}
	\end{scope}
\end{tikzpicture}
\,[\cC =\tfrac 1 2] \,.
\endpgfgraphicnamed%
\end{align}

\begin{figure}[t!]
\begin{center}%
\beginpgfgraphicnamed{graphics/temporalofflineqw}%
\begin{tikzpicture} [baseline=-2.5mm-0.5ex, x=1.5cm,y=1.5cm]
	\draw [optpath] (0,0) node [left=0pt] {$S_1$} -- +(1,1) +(0,1) node [left=0pt] {$S_2$} -- +(1,0);
	\squeezedstate{($ (0,0) + (-20:3mm) $)} {5mm} {5pt} {1pt}
	\squeezedstate{($ (0,1) + (20:3mm) $)} {5mm} {1pt} {5pt}

	\bs{(0.5,0.5)} {5mm} {0.8mm}
	\node at (0.9,0.5) {$B_1$};
	\path (0.5,0.5) +(-2.5mm,2.5mm) coordinate (B1top) +(-2.5mm,-2.5mm) coordinate (B1bottom);
	\draw (B1bottom) to [red, ->, shorten >=2pt, shorten <=2pt, bend left] (B1top);

	\draw [optpath] (1,1) -- (2,1);
	\draw [optdelay] (1,0) -- node [below=3pt] {$\Delta t$} (2,0);
	\draw [optpath] (2,0) -- +(1,1) +(0,1) -- +(1,0);

	\bs{(2.5,0.5)} {5mm} {0.8mm}
	\node at (2.9,0.5) {$B_2$};	
	\path (2.5,0.5) +(-2.5mm,2.5mm) coordinate (B2top) +(-2.5mm,-2.5mm) coordinate (B2bottom);
	\draw (B2top) to [red, ->, shorten >=2pt, shorten <=2pt, bend right] (B2bottom);
	
	\draw [fill=red] (3,0) node [right=2mm] {$D_1$} ++(-135:2mm) arc (-135:45:2mm) -- cycle;
	\draw [fill=red] (3,1) node [right=2mm] {$D_2$} ++(-45:2mm) arc (-45:135:2mm) -- cycle;
\end{tikzpicture}%
\endpgfgraphicnamed
\caption{\label{fig:temporalofflineqw}Temporal-mode GPEPS construction of a CV quantum wire using passive squeezing and linear optics.  Two single-mode squeezers~$S_1$ and~$S_2$ generate vacuum $\op p$- and $\op q$-squeezed pulses of light (respectively, as shown) at regular intervals~$\Delta t$.  These pass through a simple 50:50 beamsplitter~$B_1$, resulting in a two-mode squeezed state.  (Red arrows point from the first node to the second in Eq.~\eqref{eq:SBSpi4} for each beamsplitter.)  The delay loop in the bottom line delays the bottom mode by~$\Delta t$, allowing it to match up with the top mode of the subsequent pair emerging from~$B_1$, resulting schematically in the graph shown in Eq.~\eqref{eq:TMSqw}.  The second 50:50 beamsplitter~$B_2$ implements sequentially each of the transformations indicated by the red arrows, resulting in the final graph of Eq.~\eqref{eq:GPEPSqw}.  These pulses head toward detectors~$D_1$ and~$D_2$, which implement the necessary $\op q$-measurements (phase shifted as appropriate for the \text{\oddcolor} nodes; see Section~\ref{subsec:graphs:rules}), which are indicated by the arrows in Eq.~\eqref{eq:qwreduced}, as well as the adaptive measurement-based quantum algorithm to be implemented on the one-dimensional CV quantum wire.  The adaptiveness means that subsequent measurement bases generally must be chosen based on previous measurement outcomes.  Most measurements will involve homodyne detection in a basis that must be calculated and updated before the arrival of the next pulse, but the ability to divert the beam to an efficient photon counter is also required for universal single-mode QC~\cite{Gu2009}.}
\end{center}
\end{figure}
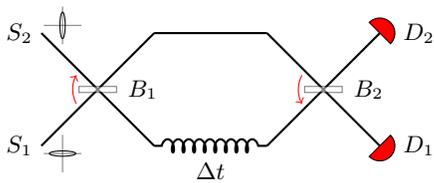

This GPEPS-based construction method has a simple and natural implementation using single-mode squeezers and linear optics.  This method uses a temporal-mode encoding~\cite{Menicucci2010}, which gives it all the advantages of the equivalent method using a single QND~gate (in terms of coherence requirements and extensibility) but improves on it because all of the required squeezing is offline.  The method is illustrated in Figure~\ref{fig:temporalofflineqw}.

A few important notes are in order.  First, periodic boundary conditions are used in the single-OPO construction~\cite{Flammia2009} to strictly enforce the requirement that~$\mat G^2 = \mat \id$, which otherwise fails at the ends.  As shown in Section~\ref{subsec:graphs:noHgraph}, however, in the construction presented here (which does not rely on exponentiating an~$\cH$-graph) these ``contaminated'' nodes can be disconnected from the rest of the cluster state by appropriate measurements of~$\op q$ on neighboring nodes, just like they are in the single-QND-gate method~\cite{Menicucci2010}.  

Second, it may not be necessary to do the projection of (the upper) half of the nodes in~$\op q$.  This was used simply to prove that the cluster state with graph~$\mat G$ can serve as a CV quantum wire.  Instead, one might be able to use the entirety of~$\mat G$ as the quantum wire, with quantum information encoded in an appropriate joint subspace of each two-node pair.  It is an open problem to determine the optimal measurements to perform in this case, but initial results indicate the possibility.  There are several reasons to think this would be useful.  First, the final cluster state in the construction above has edges with~$\cC = \frac 1 2$, while the two-mode squeezed states all had~$\cC = 1$.  While this has no effect on proofs of universality (since a non-unit edge weight simply corresponds to squeezing of the quantum information as it passes through the cluster), it could have important effects on the efficiency of the scheme with respect to the amount of initial entanglement in the two-mode squeezed states, and weights close to~$\pm 1$ in the final CV cluster-state graph are preferred~\cite{Menicucci2007}.  Also, with a joint subspace between vertical node pairs now carrying the quantum information, the other (unused) joint subspace might play an error correcting role.  Error correction and fault tolerance for CV one-way QC remains an important open problem~\cite{Ohliger2010} and will be discussed further in Section~\ref{sec:discussion}.

\section{GPEPS square-lattice cluster state}
\label{sec:GPEPSlattice}

The GPEPS techniques developed above for the CV quantum wire can be adapted to a two-dimensional square-lattice CV cluster state, as well.  This additional dimension makes the state universal for CV one-way QC~\cite{Menicucci2006}.  We again start with the states created by the single-OPO method~\cite{Menicucci2008,Flammia2009}, which have the local topology of a square lattice, but with four physical nodes per site of the lattice.  Analogous to the quantum wire, we can make this graph invariant under translations in either lattice dimension by phase shifting the highlighted nodes by~$\pi$ (which, again, is a local redefinition of basis only):
\begin{align}
\mat Z &=
\beginpgfgraphicnamed{graphics/latticeOPO}%
\begin{tikzpicture} [latticeopts]
	\def\lastnx{2}
	\def\lastnz{-2}
	\begin{scope}
	\clip (1.3,0.8,-1.9) rectangle (2,0,0.3);
	\foreach \nz in {0,...,\lastnz}
	{
	\foreach \nx in {0,...,\lastnx}
	{
		\ifthenelse {\isodd{\nz}}
			{
			\ifthenelse {\not\isodd{\nx}}
				{
				\node (ia\nx\nz) [micro-even] at (\nx,0,\nz) {};
				\node (ib\nx\nz) [micro-even] at (\nx,-1,\nz) {};
				\node (ic\nx\nz) [micro-even] at (\nx,-2,\nz) {};
				\node (id\nx\nz) [micro-even] at (\nx,-3,\nz) {};
				}
				{
				\node (ia\nx\nz) [micro-odd] at (\nx,0,\nz) {};
				\node (ib\nx\nz) [micro-odd] at (\nx,-1,\nz) {};
				\node (ic\nx\nz) [micro-odd] at (\nx,-2,\nz) {};
				\node (id\nx\nz) [micro-odd] at (\nx,-3,\nz) {};
				}
			}
			{
			\ifthenelse {\isodd{\nx}}
				{
				\node (ia\nx\nz) [micro-even] at (\nx,0,\nz) {};
				\node (ib\nx\nz) [micro-even] at (\nx,-1,\nz) {};
				\node (ic\nx\nz) [micro-even] at (\nx,-2,\nz) {};
				\node (id\nx\nz) [micro-even] at (\nx,-3,\nz) {};
				}
				{
				\node (ia\nx\nz) [micro-odd] at (\nx,0,\nz) {};
				\node (ib\nx\nz) [micro-odd] at (\nx,-1,\nz) {};
				\node (ic\nx\nz) [micro-odd] at (\nx,-2,\nz) {};
				\node (id\nx\nz) [micro-odd] at (\nx,-3,\nz) {};
				}
			}
		\node [micro-no-color] (oa\nx\nz) at (\nx+1,0,\nz) {};
		\node [micro-no-color] (ob\nx\nz) at (\nx+1,-1,\nz) {};
		\node [micro-no-color] (oc\nx\nz) at (\nx+1,-2,\nz) {};
		\node [micro-no-color] (od\nx\nz) at (\nx+1,-3,\nz) {};
		\node [micro-no-color] (za\nx\nz) at (\nx,0,\nz+1) {};
		\node [micro-no-color] (zb\nx\nz) at (\nx,-1,\nz+1) {};
		\node [micro-no-color] (zc\nx\nz) at (\nx,-2,\nz+1) {};
		\node [micro-no-color] (zd\nx\nz) at (\nx,-3,\nz+1) {};
		\ifthenelse {\not\isodd{\nz}}
			{
			\ifthenelse {\not\isodd{\nx}}
				{
				\Pitwo {ia\nx\nz} {ib\nx\nz} {ic\nx\nz} {id\nx\nz} {za\nx\nz} {zb\nx\nz} {zc\nx\nz} {zd\nx\nz}
				\Pizero {ia\nx\nz} {ib\nx\nz} {ic\nx\nz} {id\nx\nz} {oa\nx\nz} {ob\nx\nz} {oc\nx\nz} {od\nx\nz}
				}
				{
				\Pithree {ia\nx\nz} {ib\nx\nz} {ic\nx\nz} {id\nx\nz} {za\nx\nz} {zb\nx\nz} {zc\nx\nz} {zd\nx\nz}
				\Pione {ia\nx\nz} {ib\nx\nz} {ic\nx\nz} {id\nx\nz} {oa\nx\nz} {ob\nx\nz} {oc\nx\nz} {od\nx\nz}
				}
			}
			{
			\ifthenelse {\isodd{\nx}}
				{
				\Pitwo {ia\nx\nz} {ib\nx\nz} {ic\nx\nz} {id\nx\nz} {za\nx\nz} {zb\nx\nz} {zc\nx\nz} {zd\nx\nz}
				\Pizero {ia\nx\nz} {ib\nx\nz} {ic\nx\nz} {id\nx\nz} {oa\nx\nz} {ob\nx\nz} {oc\nx\nz} {od\nx\nz}
				}
				{
				\Pithree {ia\nx\nz} {ib\nx\nz} {ic\nx\nz} {id\nx\nz} {za\nx\nz} {zb\nx\nz} {zc\nx\nz} {zd\nx\nz}
				\Pione {ia\nx\nz} {ib\nx\nz} {ic\nx\nz} {id\nx\nz} {oa\nx\nz} {ob\nx\nz} {oc\nx\nz} {od\nx\nz}
				}
			}
	}
	}
	\foreach \nz in {0,...,\lastnz}
		{
		\foreach \nx in {0,...,\lastnx}
			{
			\ifthenelse {\isodd{\nz}}
				{
				\ifthenelse {\not\isodd{\nx}}
					{}
					{
					\draw [nodehighlight] (ia\nx\nz) circle (\micronodesize);
					\draw [nodehighlight] (ic\nx\nz) circle (\micronodesize);
					}
				}
				{
				\ifthenelse {\isodd{\nx}}
					{}
					{
					\draw [nodehighlight] (ia\nx\nz) circle (\micronodesize);
					\draw [nodehighlight] (ic\nx\nz) circle (\micronodesize);
					}
				}
			}
		}
	\end{scope}
\end{tikzpicture}
\endpgfgraphicnamed%
\,[\cC  = \tfrac 1 4]
\,.
\end{align}
After doing so, connections between lattice ``macronodes'' (i.e.,~groups of four nodes) look like this:
\begin{align}
\label{eq:latticetrans}
\mat Z &=
\beginpgfgraphicnamed{graphics/latticetrans}%
\begin{tikzpicture} [latticeopts]
	\def\lastnx{2}
	\def\lastnz{-2}
	\begin{scope}
	\clip (1.3,0.8,-1.9) rectangle (2,0,0.3);
	\foreach \nz in {0,...,\lastnz}
		{
		\foreach \nx in {0,...,\lastnx}
			{
			\ifthenelse {\isodd{\nz}}
				{
				\ifthenelse {\not\isodd{\nx}}
					{
					\node (ia\nx\nz) [micro-even] at (\nx,0,\nz) {};
					\node (ib\nx\nz) [micro-even] at (\nx,-1,\nz) {};
					\node (ic\nx\nz) [micro-even] at (\nx,-2,\nz) {};
					\node (id\nx\nz) [micro-even] at (\nx,-3,\nz) {};
					}
					{
					\node (ia\nx\nz) [micro-odd] at (\nx,0,\nz) {};
					\node (ib\nx\nz) [micro-odd] at (\nx,-1,\nz) {};
					\node (ic\nx\nz) [micro-odd] at (\nx,-2,\nz) {};
					\node (id\nx\nz) [micro-odd] at (\nx,-3,\nz) {};
					}
				}
				{
				\ifthenelse {\isodd{\nx}}
					{
					\node (ia\nx\nz) [micro-even] at (\nx,0,\nz) {};
					\node (ib\nx\nz) [micro-even] at (\nx,-1,\nz) {};
					\node (ic\nx\nz) [micro-even] at (\nx,-2,\nz) {};
					\node (id\nx\nz) [micro-even] at (\nx,-3,\nz) {};
					}
					{
					\node (ia\nx\nz) [micro-odd] at (\nx,0,\nz) {};
					\node (ib\nx\nz) [micro-odd] at (\nx,-1,\nz) {};
					\node (ic\nx\nz) [micro-odd] at (\nx,-2,\nz) {};
					\node (id\nx\nz) [micro-odd] at (\nx,-3,\nz) {};
					}
				}
			\node [micro-no-color] (oa\nx\nz) at (\nx+1,0,\nz) {};
			\node [micro-no-color] (ob\nx\nz) at (\nx+1,-1,\nz) {};
			\node [micro-no-color] (oc\nx\nz) at (\nx+1,-2,\nz) {};
			\node [micro-no-color] (od\nx\nz) at (\nx+1,-3,\nz) {};
			\node [micro-no-color] (za\nx\nz) at (\nx,0,\nz+1) {};
			\node [micro-no-color] (zb\nx\nz) at (\nx,-1,\nz+1) {};
			\node [micro-no-color] (zc\nx\nz) at (\nx,-2,\nz+1) {};
			\node [micro-no-color] (zd\nx\nz) at (\nx,-3,\nz+1) {};
			\BeamSplitz {ia\nx\nz} {ib\nx\nz} {ic\nx\nz} {id\nx\nz} {za\nx\nz} {zb\nx\nz} {zc\nx\nz} {zd\nx\nz}
			\BeamSplitx {ia\nx\nz} {ib\nx\nz} {ic\nx\nz} {id\nx\nz} {oa\nx\nz} {ob\nx\nz} {oc\nx\nz} {od\nx\nz}
			}
		}
	\end{scope}
\end{tikzpicture}
\endpgfgraphicnamed%
\,[\cC  = \tfrac 1 4]
\,,
\end{align}
This graph is also bipartite (with \text{\oddcolor}~and \text{\evencolor}~nodes, as shown) and satisfies~$\mat G^2 = \mat G^\tp \mat G = \mat \id$ everywhere except at a boundary.

The GPEPS construction for this state proceeds in two steps.  We start with the GPEPS construction for a grid of unattached, crisscrossing quantum wires:
\begin{align}
\label{eq:latticeTMS}
\mat Z_0 &=
\beginpgfgraphicnamed{graphics/latticeTMS}%
\begin{tikzpicture} [latticeopts]
	\def\lastnx{2}
	\def\lastnz{-2}
	\begin{scope}
	\clip (1.3,0.8,-1.9) rectangle (2,0,0.3);
	\foreach \nz in {0,...,\lastnz}
		{
		\foreach \nx in {0,...,\lastnx}
			{
			\ifthenelse {\isodd{\nz}}
				{
				\ifthenelse {\not\isodd{\nx}}
					{
					\node (ia\nx\nz) [micro-even] at (\nx,0,\nz) {};
					\node (ib\nx\nz) [micro-even] at (\nx,-1,\nz) {};
					\node (ic\nx\nz) [micro-even] at (\nx,-2,\nz) {};
					\node (id\nx\nz) [micro-even] at (\nx,-3,\nz) {};
					}
					{
					\node (ia\nx\nz) [micro-odd] at (\nx,0,\nz) {};
					\node (ib\nx\nz) [micro-odd] at (\nx,-1,\nz) {};
					\node (ic\nx\nz) [micro-odd] at (\nx,-2,\nz) {};
					\node (id\nx\nz) [micro-odd] at (\nx,-3,\nz) {};
					}
				}
				{
				\ifthenelse {\isodd{\nx}}
					{
					\node (ia\nx\nz) [micro-even] at (\nx,0,\nz) {};
					\node (ib\nx\nz) [micro-even] at (\nx,-1,\nz) {};
					\node (ic\nx\nz) [micro-even] at (\nx,-2,\nz) {};
					\node (id\nx\nz) [micro-even] at (\nx,-3,\nz) {};
					}
					{
					\node (ia\nx\nz) [micro-odd] at (\nx,0,\nz) {};
					\node (ib\nx\nz) [micro-odd] at (\nx,-1,\nz) {};
					\node (ic\nx\nz) [micro-odd] at (\nx,-2,\nz) {};
					\node (id\nx\nz) [micro-odd] at (\nx,-3,\nz) {};
					}
				}
			\node [micro-no-color] (oa\nx\nz) at (\nx+1,0,\nz) {};
			\node [micro-no-color] (ob\nx\nz) at (\nx+1,-1,\nz) {};
			\node [micro-no-color] (oc\nx\nz) at (\nx+1,-2,\nz) {};
			\node [micro-no-color] (od\nx\nz) at (\nx+1,-3,\nz) {};
			\node [micro-no-color] (za\nx\nz) at (\nx,0,\nz+1) {};
			\node [micro-no-color] (zb\nx\nz) at (\nx,-1,\nz+1) {};
			\node [micro-no-color] (zc\nx\nz) at (\nx,-2,\nz+1) {};
			\node [micro-no-color] (zd\nx\nz) at (\nx,-3,\nz+1) {};
			\tmslinkz {ia\nx\nz} {ib\nx\nz} {ic\nx\nz} {id\nx\nz} {za\nx\nz} {zb\nx\nz} {zc\nx\nz} {zd\nx\nz}
			\tmslinkx {ia\nx\nz} {ib\nx\nz} {ic\nx\nz} {id\nx\nz} {oa\nx\nz} {ob\nx\nz} {oc\nx\nz} {od\nx\nz}
			\draw [red, ->, out=-165, in=165, looseness=2] (ia\nx\nz) to (ib\nx\nz);
			\draw [red, ->, out=-15, in=15, looseness=2] (ic\nx\nz) to (id\nx\nz);
%
			}
		}
	\end{scope}
\end{tikzpicture}
\endpgfgraphicnamed%
\,[\cC = 1]\,.
\end{align}
We then apply two additional beamsplitter interactions to~$\mat Z_0$:
\begin{align}
\label{eq:latticewires}
\mat Z &=
\beginpgfgraphicnamed{graphics/latticewires}%
\begin{tikzpicture} [latticeopts]
	\def\lastnx{2}
	\def\lastnz{-2}
	\begin{scope}
	\clip (1.3,0.8,-1.9) rectangle (2,0,0.3);
	\foreach \nz in {0,...,\lastnz}
		{
		\foreach \nx in {0,...,\lastnx}
			{
			\ifthenelse {\isodd{\nz}}
				{
				\ifthenelse {\not\isodd{\nx}}
					{
					\node (ia\nx\nz) [micro-even] at (\nx,0,\nz) {};
					\node (ib\nx\nz) [micro-even] at (\nx,-1,\nz) {};
					\node (ic\nx\nz) [micro-even] at (\nx,-2,\nz) {};
					\node (id\nx\nz) [micro-even] at (\nx,-3,\nz) {};
					}
					{
					\node (ia\nx\nz) [micro-odd] at (\nx,0,\nz) {};
					\node (ib\nx\nz) [micro-odd] at (\nx,-1,\nz) {};
					\node (ic\nx\nz) [micro-odd] at (\nx,-2,\nz) {};
					\node (id\nx\nz) [micro-odd] at (\nx,-3,\nz) {};
					}
				}
				{
				\ifthenelse {\isodd{\nx}}
					{
					\node (ia\nx\nz) [micro-even] at (\nx,0,\nz) {};
					\node (ib\nx\nz) [micro-even] at (\nx,-1,\nz) {};
					\node (ic\nx\nz) [micro-even] at (\nx,-2,\nz) {};
					\node (id\nx\nz) [micro-even] at (\nx,-3,\nz) {};
					}
					{
					\node (ia\nx\nz) [micro-odd] at (\nx,0,\nz) {};
					\node (ib\nx\nz) [micro-odd] at (\nx,-1,\nz) {};
					\node (ic\nx\nz) [micro-odd] at (\nx,-2,\nz) {};
					\node (id\nx\nz) [micro-odd] at (\nx,-3,\nz) {};
					}
				}
			\node [micro-no-color] (oa\nx\nz) at (\nx+1,0,\nz) {};
			\node [micro-no-color] (ob\nx\nz) at (\nx+1,-1,\nz) {};
			\node [micro-no-color] (oc\nx\nz) at (\nx+1,-2,\nz) {};
			\node [micro-no-color] (od\nx\nz) at (\nx+1,-3,\nz) {};
			\node [micro-no-color] (za\nx\nz) at (\nx,0,\nz+1) {};
			\node [micro-no-color] (zb\nx\nz) at (\nx,-1,\nz+1) {};
			\node [micro-no-color] (zc\nx\nz) at (\nx,-2,\nz+1) {};
			\node [micro-no-color] (zd\nx\nz) at (\nx,-3,\nz+1) {};
			\beamsplitz {ia\nx\nz} {ib\nx\nz} {ic\nx\nz} {id\nx\nz} {za\nx\nz} {zb\nx\nz} {zc\nx\nz} {zd\nx\nz}
			\beamsplitx {ia\nx\nz} {ib\nx\nz} {ic\nx\nz} {id\nx\nz} {oa\nx\nz} {ob\nx\nz} {oc\nx\nz} {od\nx\nz}
			\draw [red, ->, out=-165, in=165, looseness=2] (ia\nx\nz) to (ic\nx\nz);
			\draw [red, ->, out=-15, in=15, looseness=2] (ib\nx\nz) to (id\nx\nz);
			}
		}
	\end{scope}
\end{tikzpicture}
\endpgfgraphicnamed%
\,[\cC = \tfrac 1 2]\,,
\end{align}
which results in the desired graph, shown in Eq.~\eqref{eq:latticetrans}.  This state can be projected down to an ordinary square lattice by measuring~$\op q$ on the top three nodes of each macronode (with an implicit phase shift by~$-\frac \pi 2$ on all \text{\oddcolor} nodes; see Section~\ref{subsec:graphs:rules}):
\begin{align}
\label{eq:latticeflat}
\mat Z_{\text{sl}} &=
\beginpgfgraphicnamed{graphics/latticeflat}%
\begin{tikzpicture} [latticeopts]
	\def\lastnx{2}
	\def\lastnz{-2}
	\begin{scope}
	\clip (1.3,0.8,-1.9) rectangle (2,0,0.3);
	\foreach \nz in {0,...,\lastnz}
		{
		\foreach \nx in {0,...,\lastnx}
			{
			\ifthenelse {\isodd{\nz}}
				{
				\ifthenelse {\not\isodd{\nx}}
					{
					\node (ia\nx\nz) [micro-even, \phantomfade] at (\nx,0,\nz) {};
					\node (ib\nx\nz) [micro-even, \phantomfade] at (\nx,-1,\nz) {};
					\node (ic\nx\nz) [micro-even, \phantomfade] at (\nx,-2,\nz) {};
					\node (id\nx\nz) [micro-even] at (\nx,-3,\nz) {};
					}
					{
					\node (ia\nx\nz) [micro-odd, \phantomfade] at (\nx,0,\nz) {};
					\node (ib\nx\nz) [micro-odd, \phantomfade] at (\nx,-1,\nz) {};
					\node (ic\nx\nz) [micro-odd, \phantomfade] at (\nx,-2,\nz) {};
					\node (id\nx\nz) [micro-odd] at (\nx,-3,\nz) {};
					}
				}
				{
				\ifthenelse {\isodd{\nx}}
					{
					\node (ia\nx\nz) [micro-even, \phantomfade] at (\nx,0,\nz) {};
					\node (ib\nx\nz) [micro-even, \phantomfade] at (\nx,-1,\nz) {};
					\node (ic\nx\nz) [micro-even, \phantomfade] at (\nx,-2,\nz) {};
					\node (id\nx\nz) [micro-even] at (\nx,-3,\nz) {};
					}
					{
					\node (ia\nx\nz) [micro-odd, \phantomfade] at (\nx,0,\nz) {};
					\node (ib\nx\nz) [micro-odd, \phantomfade] at (\nx,-1,\nz) {};
					\node (ic\nx\nz) [micro-odd, \phantomfade] at (\nx,-2,\nz) {};
					\node (id\nx\nz) [micro-odd] at (\nx,-3,\nz) {};
					}
				}
			\node [micro-no-color] (oa\nx\nz) at (\nx+1,0,\nz) {};
			\node [micro-no-color] (ob\nx\nz) at (\nx+1,-1,\nz) {};
			\node [micro-no-color] (oc\nx\nz) at (\nx+1,-2,\nz) {};
			\node [micro-no-color] (od\nx\nz) at (\nx+1,-3,\nz) {};
			\node [micro-no-color] (za\nx\nz) at (\nx,0,\nz+1) {};
			\node [micro-no-color] (zb\nx\nz) at (\nx,-1,\nz+1) {};
			\node [micro-no-color] (zc\nx\nz) at (\nx,-2,\nz+1) {};
			\node [micro-no-color] (zd\nx\nz) at (\nx,-3,\nz+1) {};
			\linearlinklattice {ia\nx\nz} {ib\nx\nz} {ic\nx\nz} {id\nx\nz} {za\nx\nz} {zb\nx\nz} {zc\nx\nz} {zd\nx\nz}
			\linearlinklattice {ia\nx\nz} {ib\nx\nz} {ic\nx\nz} {id\nx\nz} {oa\nx\nz} {ob\nx\nz} {oc\nx\nz} {od\nx\nz}
%
%
			}
		}
	\end{scope}
\end{tikzpicture}
\endpgfgraphicnamed%
\,[\cC = \tfrac 1 4]\,.
\end{align}
Just like for the quantum wire, this projection is useful for proving universality of the resulting state, but it may not strictly be necessary.  Instead, it may be possible to manipulate quantum information encoded within a macronode as a whole (but still using only local measurements) and\slash or to use the additional connections for error correction.  Work in this direction is ongoing.

\begin{figure}
	\def\micronodesizeorig{\micronodesize}
	\def\micronodesize{3pt}
	\def\edgethicknessorig{\edgethickness}
	\def\edgethickness{thick}
	\newcount\lastn
	\newcount\firstn
	\newcount\mval
	\newcount\lastnminone
	\newcount\lastnminm
	\newcount\nloc
	\newcount\hnext
	\newcount\vnext
	\newcount\mindex
	\newcount\vertslot
\begin{center}
\includegraphics{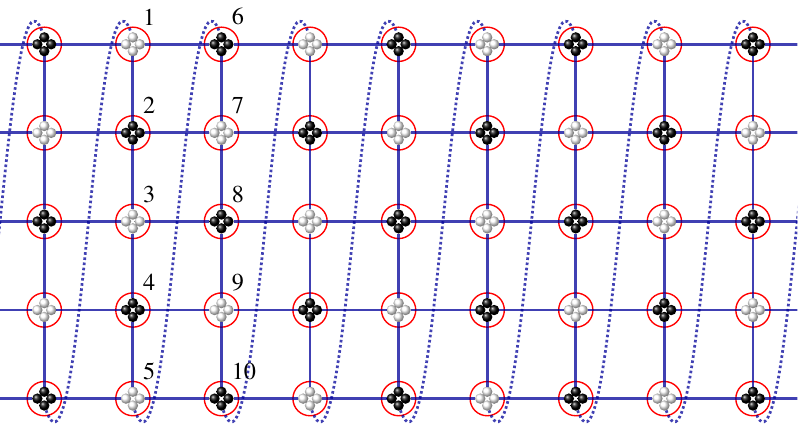}
\includegraphics{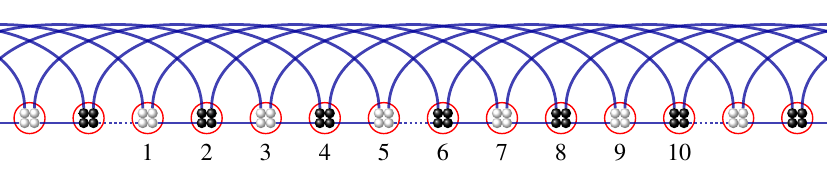}
\caption{\label{fig:2dQNDlinearized}Two equivalent representations of the GPEPS topology of the temporal-mode square-lattice CV cluster state illustrated in Eq.~\eqref{eq:latticetrans}, adapted from Figure~3 in Ref.~\cite{Menicucci2010}.  Two-mode squeezed states are arranged as in Eq.~\eqref{eq:latticeTMS}, while red circles represent the four beamsplitter transformations indicated in Eqs.~\eqref{eq:latticeTMS} and \eqref{eq:latticewires} for each macronode, as well as the projection using $\op q$~measurements on three of the micronodes within each, eventually resulting in the graph of Eq.~\eqref{eq:latticeflat}.  The two-dimensional square-lattice graph (top) is formally infinite in one dimension but finite in the other.  This graph can be redrawn as a multiply threaded infinite line graph (bottom).  There are $M$~additional threadings of the line graph that pass through macronodes $M$~units apart, resulting in a square lattice with vertical dimension~$M$.  Note that $M$~must be odd to ensure the graph is bipartite.  (As shown, ${M=5}$.)  The dotted links represent additional edges that would make the linear version translationally symmetric and equivalent to a square lattice on a cylinder with one unit of shear in the longitudinal direction.  Such a family of graphs would still be universal for one-way QC because we can measure~$\op q$ on every $M$th~node to delete it (and its links) and ``unfold'' the graph into an ordinary square lattice with a vertical dimension of \mbox{$(M-1)$}~\cite{Gu2009}.}
\end{center}
	\def\micronodesize{\micronodesizeorig}
	\def\edgethickness{\edgethicknessorig}
\end{figure}

\begin{figure}[t!]
\begin{center}
\beginpgfgraphicnamed{graphics/temporalofflinelattice}%
\begin{tikzpicture} [x=1.4cm,y=1.4cm]
	
	\draw [optpath] (0,2) node [left=0pt] {$S_1$} -- +(1,1);
	\draw [optpath] (0,2) +(0,1) node [left=0pt] {$S_2$} -- +(1,0);
	\squeezedstate{($ (0,2) + (-20:3mm) $)} {5mm} {5pt} {1pt}
	\squeezedstate{($ (0,3) + (20:3mm) $)} {5mm} {1pt} {5pt}

	\bs{(0.5,2.5)} {5mm} {0.8mm}
	\node at (0.9,2.5) {$B_1$};
	\path (0.5,2.5) +(-2.5mm,2.5mm) coordinate (B1top) +(-2.5mm,-2.5mm) coordinate (B1bottom);
	\draw (B1bottom) to [red, ->, shorten >=2pt, shorten <=2pt, bend left] (B1top);

	\draw [optpath] (1,3) -- +(1,0);
	\draw [optdelay] (1,2) -- node [below=2pt] {$\Delta t$} +(1,0);
	\draw [optpath] (2,2) -- +(1.25,1.25) -- +(3.25,-0.75);
	\draw [optpath] (2,2) +(0,1) -- +(1.75,-0.75);
	
	\bs{(2.5,2.5)} {5mm} {0.8mm}
	\node at (2.9,2.5) {$B_3$};
	\path (2.5,2.5) +(-2.5mm,2.5mm) coordinate (B3top) +(-2.5mm,-2.5mm) coordinate (B3bottom);
	\draw (B3top) to [red, ->, shorten >=2pt, shorten <=2pt, bend right] (B3bottom);
	

	\draw [optpath] (0,0) node [left=0pt] {$S_4$} -- +(1,1);
	\draw [optpath] (0,0) +(0,1) node [left=0pt] {$S_3$} -- +(1,0);
	\squeezedstate{($ (0,0) + (-20:3mm) $)} {5mm} {1pt} {5pt}
	\squeezedstate{($ (0,1) + (20:3mm) $)} {5mm} {5pt} {1pt}
	
	\bs{(0.5,0.5)} {5mm} {0.8mm}
	\node at (0.9,0.5) {$B_2$};
	\path (0.5,0.5) +(-2.5mm,2.5mm) coordinate (B2top) +(-2.5mm,-2.5mm) coordinate (B2bottom);
	\draw (B2top) to [red, ->, shorten >=2pt, shorten <=2pt, bend right] (B2bottom);
	
	\draw [optdelaylong] (1,1) -- node [below=2pt] {$M \Delta t$} +(1,0);
	\draw [optpath] (1,0) -- +(1,0);
	\draw [optpath] (2,0) -- +(1.75,1.75);
	\draw [optpath] (2,0) +(0,1) -- +(1.25,-0.25) -- +(3.25,1.75);

	\bs{(2.5,0.5)} {5mm} {0.8mm}
	\node at (2.9,0.5) {$B_4$};
	\path (2.5,0.5) +(-2.5mm,2.5mm) coordinate (B4top) +(-2.5mm,-2.5mm) coordinate (B4bottom);
	\draw (B4bottom) to [red, ->, shorten >=2pt, shorten <=2pt, bend left] (B4top);
	
	\bs{(3.5,1.5)} {5mm} {0.8mm}
	\node at (3.85,1.5) {$B_5$};
	\path (3.5,1.5) +(-2.5mm,2.5mm) coordinate (B5top) +(-2.5mm,-2.5mm) coordinate (B5bottom);
	\draw (B5top) to [red, ->, shorten >=2pt, shorten <=2pt, bend right] (B5bottom);
	
	\bs{(5,1.5)} {5mm} {0.8mm}
	\node at (5.35,1.5) {$B_6$};
	\path (5,1.5) +(-2.5mm,2.5mm) coordinate (B6top) +(-2.5mm,-2.5mm) coordinate (B6bottom);
	\draw (B6top) to [red, ->, shorten >=2pt, shorten <=2pt, bend right] (B6bottom);
	
	\draw [fill=red] (3.75,1.25) node [below right=1mm] {$D_1$} ++(-135:2mm) arc (-135:45:2mm) -- cycle;
	\draw [fill=red] (3.75,1.75) node [above right=1mm] {$D_3$} ++(-45:2mm) arc (-45:135:2mm) -- cycle;
	\draw [fill=red] (5.25,1.25) node [below right=1mm] {$D_2$} ++(-135:2mm) arc (-135:45:2mm) -- cycle;
	\draw [fill=red] (5.25,1.75) node [above right=1mm] {$D_4$} ++(-45:2mm) arc (-45:135:2mm) -- cycle;
\end{tikzpicture}%
\endpgfgraphicnamed%
\caption{\label{fig:temporalofflinelattice}Temporal-mode GPEPS construction of a square-lattice CV cluster state using passive squeezing and linear optics.  Two copies of the quantum-wire setup from Figure~\ref{fig:temporalofflineqw} are used to generate the lattice.  The upper one has the ordinary delay of~$\Delta t$ and corresponds to the vertical links in Figure~\ref{fig:2dQNDlinearized}~(top).  The longer delay of~$M\Delta t$ in the lower one gives the second threading of the wire and corresponds to the horizontal links in Figure~\ref{fig:2dQNDlinearized}~(top).  Beamsplitters~$B_3$ and~$B_4$ implement the transformations indicated by red arrows in Eq.~\eqref{eq:latticeTMS}.  (Red arrows point from the first node to the second in Eq.~\eqref{eq:SBSpi4} for each beamsplitter.)  Following this, the 50:50 beamsplitters~$B_5$ and~$B_6$ implement the transformations indicated by red arrows in Eq.~\eqref{eq:latticewires}, eventually resulting in the state with graph~$\mat Z$ from Eq.~\eqref{eq:latticetrans}.  The outputs head to four detectors, which implement the $\op q$-measurements (phase shifted as appropriate for the \text{\oddcolor} nodes; see Section~\ref{subsec:graphs:rules}) to project the state down to an ordinary square lattice, Eq.~\eqref{eq:latticeflat}, as well as the adaptive measurement-based quantum algorithm to be implemented.  The adaptiveness means that subsequent measurement bases generally must be chosen based on previous measurement outcomes.  Most measurements will involve homodyne detection in a basis that must be calculated and updated before the arrival of the next pulse, but the ability to divert the beam to an efficient photon counter is also required for universal QC~\cite{Gu2009}.}
\end{center}
\end{figure}
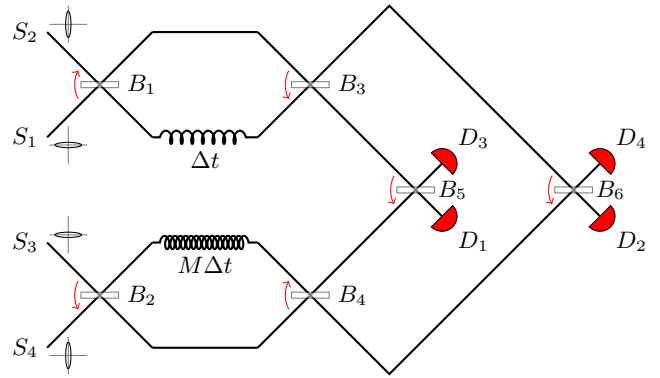

We can use the techniques from the single-QND-gate method~\cite{Menicucci2010} to design a temporal-mode encoding of this state.  The way this was done in that scheme was to ``roll up'' the square lattice onto a cylinder with a one-unit shift in the longitudinal direction.  This allows the cluster state to be generated by double threading a quantum wire.  The original threading provides the vertical links, while additional connections at a spacing of $M$~nodes apart create the second dimension of the lattice---see Figure~\ref{fig:2dQNDlinearized}.  This corresponds to a simple linear optics circuit, illustrated in Figure~\ref{fig:temporalofflinelattice}.   As shown in Section~\ref{subsec:graphs:noHgraph}, and as in the case of the quantum wire in Section~\ref{sec:GPEPSqw}, the ``contaminated'' nodes at the very beginning of the lattice can be disconnected from the rest of the cluster state by appropriate measurements of~$\op q$ on neighboring nodes, just like they are in the single-QND-gate method~\cite{Menicucci2010}.

\section{Discussion}
\label{sec:discussion}

The methods present here for generating a temporal-mode CV cluster state using offline squeezing and linear optics combine the best aspects of several proposals.  The temporal-mode encoding of the single-QND-gate method~\cite{Menicucci2010} allows finite hardware to be used repeatedly by pulsing the single-mode squeezer.  This method achieves the same result, allowing optical elements to be reused by encoding field modes in pulses of finite duration.  The advantage of this proposal is that the need for inline squeezing ($\CZ$~gate) is eliminated in favor of offline squeezing only, which is sufficient to produce two-mode squeezed states when combined with linear optics.  This was the advantage of the linear optics method~\cite{vanLoock2007}.  Finally, an extensible design combining the two was arrived at by considering the states generated in the single-OPO method~\cite{Menicucci2008}.  These states have special properties---specifically, that their graphs are bipartite and self-inverse---that allow 50:50 beamsplitters to link the two-mode squeezed pairs together.  Delay loops and phase shifts are the only other ingredients needed to generate the states.

Computing with the states is possible by measuring in~$\op q$ the extra ``layers'' of connections---which exist solely due to the use of macronodes in place of individual nodes, and three of the detectors in Figure~\ref{fig:temporalofflinelattice} can be assigned simply to this purpose.  On the other hand, it may be possible to use all four detectors in concert to perform CV one-way QC on modes that are distributed over the macronode.  The optimal measurement scheme for using these states is an open area of research.

The motivation for this is the need for error correction.  It has been known since the beginning~\cite{Menicucci2006} that finite-energy CV cluster states are, in some sense, inherently faulty due to their finitely squeezed nature.  Recent results~\cite{Ohliger2010} show this faultiness to be persistent in the sense that simple encodings of qubits or other low-dimensional systems in a finite number of CV modes cannot eliminate the errors caused by finite squeezing.  As emphasized in that paper, this does not mean that CV cluster states (made by this method or any other) are unusable for CV one-way QC.  It just means that there is no ``magic pill'' that will eliminate the faultiness introduced by finite squeezing.  Instead, error correction and fault tolerance must be addressed from the very beginning if any CV one-way QC scheme is to be scalable.  While a fault tolerant threshold~\cite{Nielsen2000} for CV one-way QC is not yet known to exist, promising results for qubit-based cluster states~\cite{Raussendorf2006} and the known ability to use CVs for fault-tolerant QC in other contexts~\cite{Lund2008} inspire confidence that such a threshold will eventually be found.

All possible schemes for QC using CV cluster states will run into this problem because the faultiness of the states is due to finite energy constraints.  The current scheme is no exception.  Nevertheless, there are several reasons to believe this is a significant improvement over other optical schemes to date.  The main reason is that the temporal-mode encoding limits the need for long-time coherence of the state since more cluster is prepared as previous pieces are consumed by the detectors.  The main theoretical limitation for the current scheme is the long delay loop in Figure~\ref{fig:temporalofflinelattice}, the length of which sets the width~$M$ of the lattice.  While computation can theoretically increase indefinitely in the horizontal dimension, the number of horizontal quantum wires linked together (and thus the number of encoded qubits or other systems) will be limited by how long this delay loop can be made without losing the ability to phaselock and modematch the interactions at the subsequent beamsplitters.  Loss and the finite coherence length of the lasers also come into play, since at least $2M$~pulses from each of the four squeezers must continue to exist coherently at all times.  Despite these restrictions, the simplicity of the setup presented in Figure~\ref{fig:temporalofflinelattice} makes it highly appealing for new experimental work.  Experiments to date, which employ the original linear optics method~\cite{vanLoock2007}, are currently limited to about four modes~\cite{Su2007,Yonezawa2010,Yukawa2008,Ukai2010}.

The main avenues for new research in this area include, first and foremost, the experimental implementation of this scheme.  Beyond that, there remain important questions about how best to adapt to the use of macronodes in place of individual nodes in the cluster.  This was alluded to several times in the text, and the issue relates both to efficiency of implementation and also to error correction, fault tolerance, and more efficient use of the squeezing resources available.  Possible extensions include higher-dimensional graphs, such as the 3D~graphs, which are useful in topological one-way QC using qubit-based cluster states and result in very high fault-tolerant thresholds~\cite{Raussendorf2006}.  The practical limitations of using only finite hardware mean that there is a maximum number of encoded qubits (i.e.,~a maximum~$M$) beyond which the scheme will cease to be feasible, if for no other reason than the coherence length of the laser is exceeded.  At this point, either the quantum wire setup from Figure~\ref{fig:temporalofflineqw} or the one for the square lattice from Figure~\ref{fig:temporalofflinelattice} will need to be concatenated with another such system in order to scale up further.  The same also applies to very long computations if we want them to be done fault tolerantly since longer computations will necessarily require larger encodings~\cite{Ohliger2010}, which reduces to the problem above.  The precise limitations and means of concatenation remain open areas of research.  Despite this, given the current state of the art in linear optics, we have good reason to believe that the method proposed here will provide an important new avenue for experimental implementation of CV one-way QC.

\acknowledgments

I thank Tim Ralph, Steve Flammia, Peter van Loock, Olivier Pfister, and Akira Furusawa for helpful discussions and suggestions.  I am also grateful to Tim Ralph, Gerard Milburn, The University of Queensland, and the ARC Centre of Excellence for Quantum Computer Technology for support during visits that contributed to this work.  Research at Perimeter Institute is supported by the Government of Canada through Industry Canada and by the Province of Ontario through the Ministry of Research \& Innovation.

%

\bibliographystyle{bibstyleNCM}
\bibliography{Temporal_offline}

\end{document}